\documentclass[sigconf, nonacm]{acmart}
\usepackage{multirow}
\usepackage{cleveref}
\usepackage{listings}
\usepackage{xcolor}
\usepackage{amsmath,amsthm}
\usepackage[most]{tcolorbox}%
\usepackage{enumitem}
\usepackage{xspace}
\usepackage[table]{xcolor} 
\definecolor{fgKeyword}{RGB}{0,92,184}   % keyword blue
\definecolor{fgRule}{RGB}{110,110,110}   % ::= and punctuation gray
\definecolor{fgFrame}{RGB}{220,220,220}  % light frame gray

\setlist{leftmargin=*}

\newtcolorbox{mybox}{
  colback=blue!5!white,
  colframe=blue!75!black,
  boxrule=0.8pt,
  arc=4pt,
  left=6pt,
  right=6pt,
  top=6pt,
  bottom=6pt,
  width=\columnwidth,
  enhanced,
}

\newcommand{\code}[1]{{\text\small\texttt{#1}}}

\newcommand{\stitle}[1]{\smallskip\noindent\textbf{#1}}

\definecolor{blue}{HTML}{5383EC}
\newcommand{\red}[1]{\textcolor{red}{#1}\xspace}
\newcommand{\blue}[1]{\textcolor{blue}{#1}\xspace}

\definecolor{orange}{RGB}{230,126,34}   % modern orange
\definecolor{teal}{RGB}{26,188,156}     % modern teal

\theoremstyle{definition} % upright text

\newtheorem{example}{Example}

\theoremstyle{plain} % italic text (typical for theorems/lemmas)

\theoremstyle{remark} % upright, smaller/less formal styling

\usepackage{listings}
\usepackage{xcolor}

\lstdefinestyle{flowguardBase}{
  basicstyle=\ttfamily\small,
  columns=fullflexible,
  keepspaces=true,
  showstringspaces=false,
  breaklines=true,
  frame=single,
  rulecolor=\color{fgFrame},
  xleftmargin=0.6em,
  xrightmargin=0.6em,
  aboveskip=0.6\baselineskip,
  belowskip=0.6\baselineskip,
  framexleftmargin=0.4em,
  framexrightmargin=0.4em,
  framextopmargin=0.35em,
  framexbottommargin=0.35em,
}

\lstdefinelanguage{flowguard}{
  sensitive=true,
  morekeywords={SOURCE,SINK,DIMENSION,CONSTRAINT,ON,FAIL,REMOVE,KILL,HUMAN,LLM,INVALIDATE,UDF},
  keywordstyle=\color{fgKeyword}\bfseries,
  morestring=[b]',
}

\lstdefinelanguage{flowguardbnf}{
  sensitive=true,
  morekeywords={SOURCE,SINK,DIMENSION,CONSTRAINT,ON,FAIL,REMOVE,KILL,HUMAN,LLM,INVALIDATE,UDF},
  keywordstyle=\color{fgKeyword}\bfseries,
  alsoletter={<>:=|,?()},
  literate=
    {::=}{{\textcolor{fgRule}{::=}}}3
    {|}{{\textcolor{fgRule}{|}}}1
    {,}{{\textcolor{fgRule}{,}}}1
    {?}{{\textcolor{fgRule}{?}}}1
    {(}{{\textcolor{fgRule}{(}}}1
    {)}{{\textcolor{fgRule}{)}}}1
    {<}{{\textcolor{fgRule}{<}}}1
    {>}{{\textcolor{fgRule}{>}}}1
    ,
}

\lstnewenvironment{FlowGuardBNF}
  {\lstset{style=flowguardBase,language=flowguardbnf}}
  {}

\lstnewenvironment{FlowGuardExample}
  {\lstset{style=flowguardBase,language=flowguard}}
  {}

\newcommand\vldbavailabilityurl{https://github.com/ElaineAng/db-fork}
\newcommand{\sys}{{B}\textsc{ranch}{B}\textsc{ench}\xspace}

\begin{document}
\title{\sys: Aligning Database Branching with Agentic Demands}

%%
%% The "author" command and its associated commands are used to define the authors and their affiliations.
\author{Elaine Ang}
\affiliation{%
  \institution{Columbia University}
}
\email{ra3448@columbia.edu}

\author{In Keun Kim}
\affiliation{%
  \institution{Columbia University}
}
\email{ik2619@columbia.edu}

\author{Sam Weldon}
\affiliation{%
  \institution{Columbia University}
}
\email{sw3927@columbia.edu}

\author{Kevin Durand}
\affiliation{%
  \institution{Columbia University}
}
\email{
  kpd2136@columbia.edu}

\author{Kostis Kaffes}
\affiliation{%
  \institution{Columbia University}
}
\email{kkaffes@cs.columbia.edu}

\author{Eugene Wu}
\affiliation{%
  \institution{Columbia University}
}
\email{ewu@cs.columbia.edu}

\begin{abstract}

Branchable databases are evolving from developer tools to infrastructure for agentic workloads characterized by speculative mutations and non-linear state exploration. Traditional RDBMS mechanisms such as nested transactions do not provide the persistent isolation and concurrent branch management required by autonomous agents, and recent ``zero-copy'' designs make different trade-offs whose impact on agentic workloads remains unclear.

To clarify this space, we present \sys, a benchmark for evaluating branchable relational DBMSes under agentic exploration. We characterize five representative workloads---agentic software engineering, failure reproduction, data curation, MCTS, and simulation---and design parameterized macrobenchmarks that execute {\it branch–mutate–evaluate} loops to reflect these workloads, along with microbenchmarks that isolate branch lifecycle costs.

We evaluate state of the art systems including Neon, DoltgreSQL, Tiger Data, Xata, and PostgreSQL baselines, and find a fundamental tension: systems optimized for fast branching suffer up to $5-4000\times$ slower reads as branches deepen, while systems optimized for fast data operations incur $25-1500\times$ higher branch creation and switching latency. Further, no current system supports the representative workloads at scale. 
These results highlight the need for branch-native DBMSes designed specifically for agentic exploration.

\end{abstract}

\maketitle

%%% ArXiv version - VLDB-specific blocks removed %%%
\pagestyle{plain}
\vspace{.3cm}
\begingroup\small\noindent\raggedright
% \textbf{Note:} Submitted to Proceedings of the VLDB Endowment (PVLDB).
% Uncomment the line below if/when accepted and update volume/issue/year:
% \textbf{Note:} To appear in Proceedings of the VLDB Endowment (PVLDB), Vol. \vldbvolume(\vldbissue), \vldbyear.
\\[0.5em]
\textbf{Code and Artifacts:} The source code, data, and artifacts are available at \url{\vldbavailabilityurl}.
\endgroup
%%% End ArXiv version %%%

\section{Introduction}
\label{sec:intro}
The recent emergence of agentic systems is shifting systems from executing a single execution plan to managing large-scale \emph{agentic speculation}.
To solve a high-level task, agents generate and execute a tree of hypothetical actions, and evaluate their effects in order to select the best results.
On Terminal-Bench~\cite{tbench}, for instance, Monte Carlo Tree Search (MCTS) can improve task success rates from $3.4\% \to 30.6\%$~\cite{xu2025toward}. These improvements stem from the ability to efficiently fork, evaluate, and discard many alternative states, rather than from better LLM reasoning.

\begin{figure}[t]
    \centering
    \includegraphics[width=.85\columnwidth]{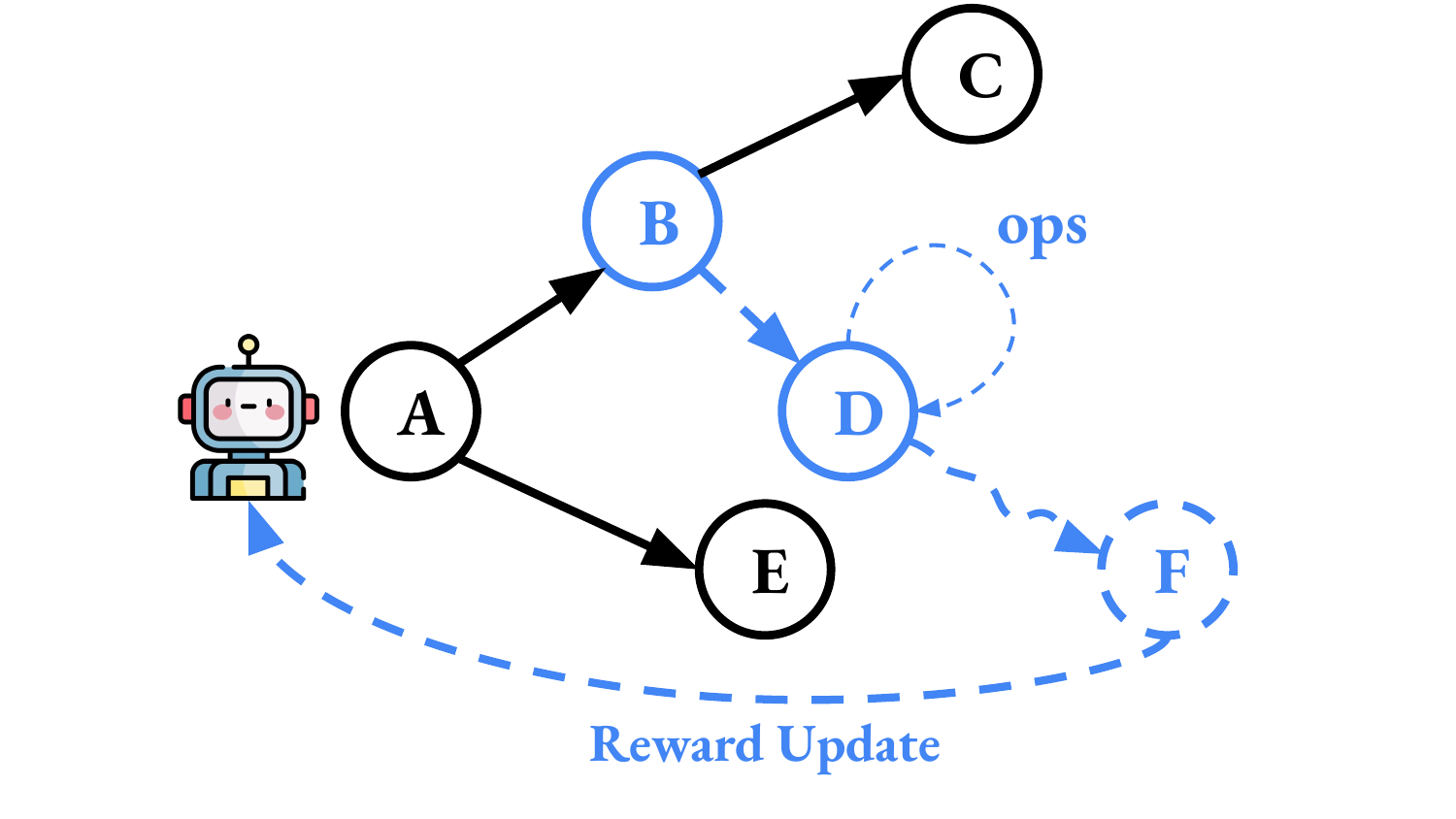}
    \caption{Agentic exploration via Monte Carlo Tree Search over database state. The agent explores a tree of hypothetical actions by forking database branches, executing operations, evaluating outcomes, and selecting optimal paths.}
    \label{fig:mcts-over-db}
    % https://docs.google.com/presentation/d/1RpBrxf1tnI_KreVEHF39Q0Ot0tf5zpif1Ba0V2cTidI/edit?slide=id.g3cac4463b03_1_12#slide=id.g3cac4463b03_1_12
\end{figure}

Agentic speculation shifts systems bottlenecks.
Traditional database systems are designed to isolate concurrent users that execute over a linear progression of shared state.
In contrast, a single agent can generate a multitude of speculative worlds.
The critical design objective is not only query latency or transaction throughput, but speculative state management: the ability to checkpoint, fork, restore, evaluate, and merge/discard database state at high frequency without prohibitive storage or performance costs.

In response, a new class of \emph{branchable DBMSes} expose an explicit \texttt{branch} abstraction: users can create a new writable branch from a current or past state, switch between branches, and compare (and in some systems, merge) divergent histories using Git-like operations.
However, existing systems implement this abstraction using fundamentally different mechanisms.
For example, Neon uses storage-layer copy-on-write by marking the appropriate LSN in the write-ahead log and sharing unchanged pages with a branch's parent~\cite{neon-intro} while
%   over \emph{log-indexed timelines} (new branches begin at a chosen WAL LSN and share unchanged pages with their parent)
Xata uses block-level copy-on-write under an unmodified PostgreSQL~\cite{xata-intro}.
%at the \emph{block storage layer}
Dolt versions the database state using a custom \emph{content-addressed Merkle/prolly-tree} storage engine, and supports structural sharing and efficient \texttt{diff}/\texttt{merge} semantics~\cite{dolt-intro}.
Other approaches approximate branching via filesystem snapshots or physical cloning in vanilla PostgreSQL, trading simplicity for coarse granularity and high overheads~\cite{pg-file-copy}.
These architectural designs  differ in copy-on-write granularity, metadata organization, and garbage collection that each target different branch-related workloads.
\S\ref{sec:taxonomy} develops a taxonomy of zero-copy branching mechanisms based on the targeted DBMS layer and their supported branching features.
%, we organize these mechanisms into a taxonomy based on (i) the DBMS layer where branching is implemented (e.g., storage/logical), and (ii) the feature surface and semantics they expose (e.g., restore-from-history, diff, merge, isolation).

To make these requirements concrete, we consider five representative agentic applications that span the structural space of speculative branching and exercise the typical \emph{branch–mutate–evaluate} loop.  We use MCTS as a representative example:

\begin{example}\it
    \Cref{fig:mcts-over-db} illustrates MCTS over a branchable DBMS. Each node in the search tree corresponds to a database branch representing a hypothetical world state. In one iteration, the agent selects a branch, forks a child, applies a short sequence of mutations (e.g., \code{ops}), and evaluates the resulting state---either via internal queries, cross-branch comparisons, or external/LLM scoring. Even for a single task, this loop can generate thousands of branches with lifetimes ranging from milliseconds (failed rollouts) to minutes (promising subtrees), stressing branch lifecycle latency, metadata scalability, garbage collection, and cross-branch reads.
\end{example}

Beyond MCTS, we consider four additional workloads to stress different dimensions of branching: \emph{agentic software engineering}, \emph{failure reproduction}, \emph{data curation}, and \emph{simulation}.
Software engineering and failure reproduction emphasize rapid branch creation and high-throughput execution. Data curation stresses cross-branch analytics and comparison. Simulation generates wide bursts of short-lived branches. These diverse applications span depth, fanout, mutation intensity (logical and physical), and branch life-cycle management.  These conflicting requirements stress-test branch creation cost, storage amplification, execution efficiency, and isolation.

These pressures highlight four key desiderata for branchable DBMSes:
(i) fast and resource-efficient branch creation, switching, and deletion;
(ii) scalability to deep and wide branch trees with efficient branch management and garbage collection;
(iii) efficient execution of hybrid transactional and analytical workloads over branched state, including cross-branch comparisons; and
(iv) performance isolation so that speculative branches do not unpredictably perturb the base state or other branches.

These desiderata are in tension. Mechanisms optimized for fast metadata-only branching may degrade under heavy physical mutations; systems tuned for deep lineage retention may incur high overhead under bursty fanouts; aggressive reclamation can conflict with flexible merging or comparison. Understanding these trade-offs—and determining which architectures are best suited to which workload classes—requires a benchmark that systematically varies exploration structure and intensity.

Unfortunately, contemporary benchmarks fundamentally assume a ``single branch'' setting and do not measure fork latency, branch tree scalability, cross-branch comparison cost, or garbage collection under high branching loads. Motivated by SEQUOIA 2000~\cite{novel-database-benchmark}, which exposed mismatches between new scientific workloads (e.g., spatial data processing, earth observation queries) and existing benchmarks, we argue that branch-intensive agentic workloads demands a new class of benchmarks.

%Software engineering and failure reproduction emphasize fast branch creation/deletion, and high-throughput query execution to run tests and replay history.
%Data curation explores alternative cleaning and transformation pipelines and stresses cross-branch analytics.
%MCTS stresses repeated branch-mutate-evaluate loops along deep dependency chains.
%Simulation generates wide bursts of extremely short-lived branches that require fast branch creation and scalable bookkeeping.
%This diversity makes it difficult to reason about branching design trade-offs---or to compare systems fairly---without a benchmark that can systematically vary exploration shape, branch lifetime, and mutation intensity.

%These workloads suggest that branchable DBMSes must satisfy a broad set of desiderata:
%(i) low-latency branch creation, connection, and deletion;
%(ii) scalability to deep and wide branch trees (including efficient garbage collection as branches proliferate and expire);
%(iii) efficient execution of hybrid transactional and analytical work over branched state, including cross-branch comparisons and aggregations;
%and (iv) a degree of performance isolation so speculative branches do not unpredictably perturb the latency or resource usage of other branches or the base state.
%A useful benchmark must therefore specify not only a query mix, but also branching frequency, tree shape (depth/width), branch lifetime distributions, cross-branch read/compare intensity, and parallel exploration.

To address this gap, we introduce \sys, a benchmark to evaluate branchable relational DBMSes under an agent's exploration patterns. \sys includes (1) \emph{macrobenchmarks} that instantiate parameterized \emph{branch–mutate–evaluate–compare} loops that reflect the above workloads, and (2) \emph{microbenchmarks} that isolate branch lifecycle primitives to separate mechanism overheads from query execution. We use \sys to evaluate representative systems across branching architectures, including storage-layer approaches (Neon, Xata, Tiger), Merkle/prolly-tree versioning (Dolt), and PostgreSQL baselines.  In general, current systems exhibit:
\begin{itemize} [leftmargin=*]
    \item \stitle{Limited Capabilities.} No system fully supports all agentic workflows at scale. PostgreSQL-native baselines (transactions/savepoints, file-copy clones)  are only able to execute a very narrow subset of the workflows we consider.
    \item \stitle{Performance Trade-offs.} When comparing the two most feature-complete systems, Neon and Dolt, Neon's data operations incur near-zero overhead, but Neon is also $25\times$ ($1500\times$) slower to create (switch) branches compared to Dolt. In contrast, Dolt's branch operations are near-instant, but reads are $5-4000\times$ slower as branches and mutations accumulate.
\end{itemize}
Our results demonstrate that existing branchable databases optimize for either branching agility or query performance, but not both, leaving a critical gap for systems designed to support high-frequency speculative exploration over persistent, mutable state.

\smallskip
\noindent In addition, we contribute:

\begin{itemize}[leftmargin=*]
    \item \textbf{Workload characterization.}
          We characterize branching demands in agentic speculation via five representative workflows---agentic software engineering, failure reproduction, data curation, MCTS, and simulation---and distill their requirements into executable branch and workload patterns (\S\ref{sec:workload}).

    \item \textbf{Branchable DBMS taxonomy.}
          We develop a taxonomy that categorizes systems by where branching is implemented in the stack and by the branch operations and semantics they support (e.g. diff, merge, isolation) (\S\ref{sec:taxonomy}).
    \item \textbf{Benchmark suite.}
          We design \sys, a benchmark suite that combines (i) macrobenchmarks for the five workflows above with (ii) microbenchmarks that isolate branch lifecycle costs (\S\ref{sec:benchmark}). Through an extensive empirical study, we provide a decision framework for aligning branching strategies with agentic demands and highlight areas for future research and optimization (\S\ref{sec:evaluation}).
\end{itemize}
\section{The need for branching}
\label{sec:workload}
Modern agents do not merely read from databases. They explore a space of hypotheses by instantiating new states---extracted ML features, schema mutations, temporary indexes, and updated data---to evaluate downstream objectives. This iterative exploration is inherently divergent: agents fork state into multiple parallel trajectories to compare outcomes, rapidly discard suboptimal branches, and merge ``winning'' states back.   We refer to this distinctive access pattern as \textbf{Agentic Speculative Branching}. With \emph{branch}, we denote a named, isolated, and independently mutable database state derived from a parent state, where the system {\it shares} as much of the unchanged data across the branches as possible.

While existing database engines support branching through mechanisms such as transaction savepoints, physical table copies, or temporary tables, these techniques were designed for occasional branches in traditional applications. In contrast, we anticipate that future agentic workloads will generate branch operations at such scale and frequency that current designs incur prohibitive performance and storage overheads.

This section describes five workflows that each generate sustained, high-frequency branching workloads that stress latency, storage, and branching semantics. In these settings, the database ceases to be a passive storage layer and instead becomes the agent’s working memory and experimental substrate. We use these workflows to motivate requirements of a branchable database and the design of our benchmarking suite.

\stitle{Scope.}
We focus on the DBMS workload generated by {\it single-task exploration}: all branches are derived from one root state, and there are no concurrent users or other agentic tasks accessing the database. This scoping decision isolates the challenge of managing exploration trees without conflating it with traditional multi-user concurrency control. Even in this simplified setting, we find that existing branchable database designs are pushed to their limits (\S~\ref{sec:evaluation}), motivating our focus on single-agent performance before addressing multi-agent coordination.

%This shift in interaction model includes five representative workflows: agent assisted software development (e.g., vibe coding), diagnostic and failure reproduction (e.g., root-cause analysis on production snapshots), data cleaning, Monte Carlo Tree Search, and Monte Carlo simulation.

%We argue that in order to serving these workflows efficiently, the core challenge lies in the structural mismatch between the workload generated by these workflows and existing database engines. An agent’s trajectory consists of high-frequency writes, frequent schema changes, zero-copy clones, and rapid rollbacks—operations that traditional mechanisms, such as physical table copies, temporary tables, or transaction savepoints, cannot support without prohibitive performance or storage overhead. Serving these workflows thus requires an architecture capable of near-instantaneous branching and native versioned state management. In the remainder of this section, we characterize the unique system demands of each pillar, establishing the requirements for our benchmarking suite.

\subsection{Representative Workflows}
\label{sec:workload:workflows}

To stay consistent with our benchmark, our motivating scenarios build on TPC-H and TPC-C schemas.

\subsubsection{Agentic Software Engineering (ASE)}
Today's ASE already runs dozens of agents in parallel to fix bugs, develop features, refactor code, and redesign the application.  Each agent makes
%In agent-assisted development (``vibe coding''), a developer instructs an LLM
%agent to build application features that require 
iterative changes to the codebase and database~\cite{replit-database, electric-sql-sandbox, replit-migrations, kimi-k2.5, neon-branching-workflows}.
An iteration may involve reading the current database, apply schema changes or migrations
(\texttt{ALTER TABLE}, \texttt{CREATE INDEX}), backfill and update data, and run
the application's tests that execute many read-heavy queries.
Each iteration is speculative until all tests pass, and must be rolled back if any tests fail.

Recent work has shown that when multiple software engineering agents run in parallel, uncoordinated changes make the system brittle~\cite{cursor-parallel-agents}.
Thus, branching is both used to isolate concurrently developed features from each other, and to checkpoint state-mutating actions so the agent can easily revert.

%Branching thus serves two purposes: isolating concurrent features from one another, and checkpointing individual state-mutating steps within a feature so the agent can cheaply revert to the state before a destructive step rather than undoing it in place.

%The tree's width reflects the number of concurrent features or alternative strategies, while its depth reflects the number of checkpointed steps within a feature. Beyond schema changes and data mutations, each branch also executes read-heavy test queries to validate the new state.

%Each such migration is speculative: its correctness is unknown until tests pass, and rolling back a failed migration on a live database is error-prone, since reverting a schema change or backfill may cascade into downstream indexes, constraints, or materialized views that must also be reverted.

\begin{example}\it
  Consider an agent implementing a new feature: adding  customer loyalty tiering.
  The agent forks \texttt{B1} from \texttt{main} before adding a
  \texttt{tier} column to \texttt{customer}, then forks \texttt{B2} from \texttt{B1} before backfilling \texttt{tier} based on customer payment histories.
  But test queries show almost all customers are assigned to the lowest tier due to poor tiering thresholds.
  The agent forks \texttt{B3} from \texttt{B1} to revert the backfill action, then backfills with new thresholds, and runs the tests again.
  The tests pass, so \texttt{B3} is merged to \texttt{main} and the others are deleted.
  %Concurrently, a second agent develops a spending-limit feature on its own branch, also adding a column to \texttt{customer}; because both branches modify the same base table, their changes remain invisible to each other and to live queries on \texttt{main} until merge.
\end{example}

\subsubsection{Failure Reproduction}
An error in production  may be due to a past database transaction that lay dormant and then triggered the error.   The goal of failure reproduction is to reproduce the error and identify the transaction that triggered it.

Existing methods replay transaction history of size $N$~\cite{trod,r3}, which could benefit from simple binary search:
%
%After a failure has been mitigated in production, the root cause often remains unknown.
%The agent's goal is to reproduce the failure on a snapshot of pre-failure state and pinpoint which transaction triggered it~\cite{trod, r3}.
%given the transaction log, the agent performs a binary search:
the agent forks from a known-good database state in the past, replays a
prefix of the log, and runs tests to trigger the error. % checks whether the corruption is present.
By comparing branches with and without the error, the agent
narrows the window until it isolates the culprit transaction.

In expectation, the agent creates $log_2(N)$ branches from the known-good state, which produces a wide and flat exploration tree.     Each branch is write-intensive: it replays a prefix of transactions, then executes read queries to check invariants and test the database state.
%All branches fork from the same root snapshot, producing a wide, flat
%tree whose width grows as the search narrows.
%The workload on each branch is write-intensive---replaying transaction sequences via \texttt{INSERT}, \texttt{UPDATE}, and occasional schema changes---followed by read queries that check invariants.
Since the agent rapidly creates and discards branches, the system must provide fast branch creation and reset, and high write-throughput for replaying transactions.
%, and provenance of each branch's mutation history for auditability.

\begin{example}\it
  A test fails when attribute \texttt{o\_ol\_cnt} in \texttt{orders} does not match the actual row count in \texttt{order\_line} for commodities orders, and the agent must analyze a log of 1000 transactions.
  It forks and replays the first 500 transactions, and finds that the constraint fails.
  It discards the branch, forks,  replays and checks the first 250 transactions, and so on until it isolates transaction 372.
  The final branch's mutation history further serves as a reproducible test case.
  %It forks a branch from the last known-good snapshot, replays the
  %first 500 transactions, and runs the count-mismatch check---corruption present, so the branch is discarded. It forks again, replays only 250---clean, discarded. It forks with 375---corruption, discarded. Continuing this binary search one branch at a time, the agent converges on transaction~\#347---a delete from the nightly cleanup job that removed order lines belonging to an in-flight order. 
\end{example}

%\ewu{What dimensions will get pushed?}

\subsubsection{Data Curation}
A major part of data curation is the process of detecting and correcting errors in a database to improve its quality for downstream use~\cite{data-cleaning}.
There is an increasing shift towards continuous, online data quality monitoring of data corpora~\cite{schelter2018automating, data-quality-online}, as well as search-based methods to find sequences of cleaning operations to address detected errors~\cite{saga, alphaclean, cleanml, cleaning-for-ml}.

It is thus reasonable to anticipate a future where a pool of agents continuously identify potential data errors in subsets of tables, and then spawn additional agents to clean those errors.
This exploration pattern creates one branch to clean each detected anomaly---perhaps dozens concurrently. Agents clean an anomaly by exploring paths of cleaning operations; each operation follows a branch-mutate-validate pattern to explore different operations and hyperparameters (e.g., thresholds, window sizes).   Mutations may create intermediates or update/delete tuples, while validation may read data in the current branch or perform cross-branch reads to compare outcomes across candidate cleaning paths.
To summarize, this workload demands high write-throughput and efficient scans per
branch, and fast cross-branch reads.

%Running them sequentially on a single copy risks one strategy masking or conflicting with another's corrections; comparing their effects requires either expensive undo or redundant copies of the data. Branching provides a natural solution: each strategy runs on its own branch, applying data mutations (bulk \texttt{UPDATE}s, \texttt{DELETE}s, or \texttt{CREATE TABLE \ldots\ AS SELECT}) to its isolated copy of the dataset.
%The agent validates progress with intra-branch reads (data-quality
%checks on the branch itself) and inter-branch reads (comparing
%outcomes across branches to identify the most effective strategy).
%Each strategy may also have tunable hyperparameters (e.g.\ threshold
%values, window sizes, or confidence levels), so the agent can fork
%additional branches to sweep over configurations within a strategy.
%The total number of branches scales with both the number of strategies
%and the granularity of hyperparameter search, producing a wide, shallow
%tree.

%Real-world datasets require multiple cleaning strategies---e.g.\ constraint-based repair, statistical imputation, outlier removal---that often operate on overlapping regions of the same tables.

\begin{example}\it
  An agent detects that the \texttt{customer} table has missing \texttt{c\_balance} values and suspicious outliers in \texttt{c\_ytd\_payment}.  It creates one exploration subtree try different imputation strategies, one branch to drop nulls, and a third subtree to clip outliers at varying percentile thresholds.     After each cleaning operation, the agent runs data-quality checks and compares across the branches to discard poor cleaning candidates and select the final outcome.
  %It forks three branches, one per strategy: the first imputes missing
  %balances with the district average, the second drops rows with nulls
  %entirely, and the third caps outliers at a percentile threshold.
  %The third strategy has a tunable cutoff, so the agent forks two
  %sub-branches to try the 95th and 99th percentiles.
  %On each branch the agent runs data-quality checks (null rate, standard
  %deviation) and compares results across branches to select the best
  %outcome.
  %The winning branch is merged to \texttt{main}; the rest are discarded.

\end{example}

\subsubsection{Monte Carlo Tree Search}
Monte Carlo Tree Search (MCTS) is an optimization and search algorithm that is increasingly applied to database-backed problems such as
financial and supply chain management~\cite{mcts-trading, mcts-inventory},
text-to-sql~\cite{mcts-sql}, query optimization~\cite{learned-rewrite},
and LLM post-training workflows ~\cite{mcts-exact}.

The algorithm selects a promising leaf, expands it with an untried action, simulates a random rollout to estimate value, and backpropagates the result~\cite{uct, mcts-survey}.
Each iteration applies an \texttt{UPDATE}/\texttt{INSERT} during expansion to create a new
state, and an analytical evaluation query after a rollout to estimate a score.
The resulting branch tree incrementally grows to become deep and narrow---successive refinements can have tens or hundreds of levels, while each state may have $3-10$ children.
MCTS is often parallelized to expand different subtrees concurrently, and to manage storage, a background process prunes low-value branches.

%Multiple threads often expand different regions of the tree concurrently,
%so the system must support low branch-creation latency, stable analytical
%query performance as branch cardinality grows into the hundreds, and
%efficient garbage collection to reclaim storage as low-reward subtrees are
%pruned.

\begin{example}\it
  An agent plans which warehouse will fulfill each pending order to minimize total shipping cost. Each tree level corresponds to one order; each branch at that level assigns the order to a different warehouse. Expansion forks a new speculative branch and applies the assignment via an UPDATE (e.g., \texttt{UPDATE stock SET (...)}) to reduce that warehouse’s inventory within the branch’s state and constrain the viable warehouses for subsequent orders.

  Rollout assigns the remaining orders greedily and scores the complete
  plan with an analytical query such as \texttt{SELECT SUM(ol\_amount) FROM order\_line JOIN warehouse}.
  Rewards are backpropagated to favor low-cost assignment sequences.
  The selection policy deprioritizes subtrees whose early assignments
  led to stockouts or high final costs.  Search terminates after a fixed iteration budget, and the
  highest-reward complete path is adopted as the fulfillment plan.

\end{example}

\subsubsection{Simulation}
Monte Carlo simulation estimates an outcome distribution by running many independent trials with randomized inputs and aggregating the results~\cite{why-mcs}.
For database applications such as loan portfolios, inventory networks, or patient cohorts~\cite{simsql, databricks-mc-var, ketteq-neon, mcs-risk, mcs-healthcare}, each trial executes tens or hundreds of \texttt{INSERT} or \texttt{UPDATE} transactions to simulate future possibilities.  This results in a wide, flat tree where  the initial state is forked thousands of times, each representing a sample trial. Each trial runs independently, so the system must sustain high aggregate write throughput and efficient batch branch creation to ensure branching does not dominate runtime.
%While individual operation latency is secondary, 
The database must provide fast cross-branch aggregation queries (e.g., percentiles or expected shortfall) to synthesize the results of the entire simulation, as well as efficient pruning and garbage collection to discard ephemeral branches.

\begin{example}\it
  A supply-chain risk agent assesses inventory shortfall by forking 1000 branches
  to simulate 30 days of possible order fulfillment patterns in each branch.
  Each simulated day inserts into \texttt{orders}, updates \texttt{stock}, and triggers replenishment when \texttt{stock.s\_quantity} for an item falls below a threshold.
  After all trials complete,  the agent runs a single cross-branch aggregation to calculate the stockout frequency and total fulfillment cost across all branches and reports the distribution.  It then discards all branches.
\end{example}

\subsection{Requirements for Branchable Databases}
\label{sec:workload:characteristics}

The workflows highlight several interrelated requirements for branchable databases that a benchmark should exercise.
%The five workflows impose a common set of requirements on the underlying branchable database.

\stitle{{Fast Branch Management}.}
Branching is the primary control structure, so branch creation and deletion must be fast to not bottleneck per-iteration latencies, particularly under concurrent workloads.
%Every workflow creates and discards branches as its primary control structure. Branch creation and deletion must be fast enough that they do not dominate per-round cost, even when branches are created concurrently by multiple threads or discarded in rapid succession.

\stitle{{Flexible Tree Topology.}}
The scenarios exhibit topologies that vary across
flat stars (simulation, failure reproduction),
wide shallow trees (data cleaning), moderate bushy trees (software
development), and deep narrow trees (MCTS).
The system must efficiently manage branch metadata---parent-child
lineage, visibility propagation, and per-branch state---and maintain
stable performance as both depth and breadth grow.

\stitle{Schema \& Data Mutations.}
The database must handle varying mixtures of schema and data mutations without degrading branching and query performance:
software development runs schema migratios and data backfills;
failure reproduction replays recorded transaction logs with few schema changes;
data cleaning performs bulk and tuple-level table rewrites:
MCTS updates a few records at a time; and
simulation runs heavy data mutation per branch with almost no schema changes.

\stitle{Interleaved read-heavy evaluation.}
The effects of mutations are evaluated with read queries---integration tests, invariant checks, data-quality metrics, reward queries, or outcome aggregations.
These must execute robustly, even at  high branch cardinality.

\stitle{Cross-branch queries.}
Agents regularly read and compare database states across branches to e.g., select the best curation pipeline, outcome distributions after simulation, or compute the diff between two schema versions.
%At the end of a workflow, diffing or aggregation queries read across surviving branches to compare or summarize outcomes---e.g.\ selecting the best curation pipeline or computing simulation percentiles, and compare the diffs between two schema versions. These queries must efficiently access data from many branches.

\stitle{Logical and Performance Isolation.}
%Most branches differ by small deltas relative to a shared base, so efficient branching demands copy-on-write versioning that shares unchanged data while isolating each branch's modifications.
Branches must be logically isolated to prevent effects in one branch from affecting others.
Branches should also exhibit performance isolation so that heavy workloads on one branch do not degrade the performance on others.

%: schema change on a feature branch must not disrupt \texttt{main}, and destructive replay must not propagate to the production snapshot.
%Performance isolation prevents heavy workloads on one branch from degrading latency or throughput on others.

\section{Branchable Database Designs}
\label{sec:taxonomy}
The naive approach to branching is do a full dump/restore, or to make a copy of the database (\texttt{CREATE DATABASE ... TEMPLATE ...}). Unfortunately, dump/restore does not {\it share} overlapping data between the branches and quickly becomes prohibitively expensive. Even though PostgreSQL 18 supports creating a new database with \texttt{file\_copy\_method=clone}, that requires Copy-on-Write filesystem support~\cite{pg-file-copy}, and has strict requirements on concurrent operations on the database to prevent data inconsistency, making that unsuitable for more demanding agentic speculative workflows (\S\ref{sec:workload}).

This section first describes why existing nested transaction support is not sufficient for agentic database branching, and then organizes existing {\it zero-copy branching} mechanisms into a taxonomy based on the DBMS layer where the copy-on-write (CoW) logic is applied (\Cref{tab:taxonomy}).
Our focus is on use cases that involve large rates of branching operations, as well as data and schema mutations.  As such, we focus on transactional and HTAP systems, and discuss branchable OLAP systems in \Cref{sec:related}.

%The goal is to maximize the amount of data {\it sharing} with low latency.  
%Despite exposing different branching APIs (snapshots, clones, forks, sandboxes), most branchable databases follow a handful of designs based on the layer where copy-on-write (CoW) is performed. 

% \begin{figure}
%     \centering
%     \includegraphics[width=\linewidth]{figures/categories.png}
%     \caption{Taxonomy of different branching mechanisms, organized by DBMS layer. \ewu{not a fan of the naming.  Should include example systems in figure -- maybe on the left side}  }
%     \label{fig:categories}
%     % slides:
%     % https://docs.google.com/presentation/d/1RpBrxf1tnI_KreVEHF39Q0Ot0tf5zpif1Ba0V2cTidI/edit?usp=sharing
% \end{figure}

\begin{table}
    \centering
    \footnotesize
    \begin{tabular}{p{1.8cm} p{2.3cm} p{3.4cm}}
        \arrayrulecolor{gray!50}
        \toprule
        \textbf{DBMS Layer} & \textbf{Branch Mechanism} & \textbf{Systems}                    \\
        \midrule
        Storage Substrate   & Block-level CoW           & TigerData, Xata, Vela, PG file copy \\
        \midrule
        Recovery            & WAL-based                 & Neon                                \\
        Manager             & reconstruction            &                                     \\
        \midrule
        \multirow{3}{1.8cm}{Storage Manager}
                            & Content-addr. tree        & Dolt                                \\
                            & Page-level CoW            & Minuet (research system)            \\
                            & Delta overlays            & Decibel (research system)           \\
        \bottomrule
    \end{tabular}
    \caption{Taxonomy of branchable database designs. }
    \label{tab:taxonomy}
\end{table}

\begin{table*}[t]
    \centering
    \small
    \begin{tabular}{l c c c c  c c c  c c  c c}
        \arrayrulecolor{gray!50}
                        & \multicolumn{4}{c}{\textbf{Single Branch}} & \multicolumn{3}{c}{\textbf{Cross Branch}} & \multicolumn{2}{c}{\textbf{Branch Meta}} & \multicolumn{2}{c}{\textbf{Data and Schema}}                                                                                            \\
        \cmidrule(lr){2-5} \cmidrule(lr){6-8} \cmidrule(lr){9-10} \cmidrule(lr){11-12}
        \textbf{System} & \textbf{Create}                            & \textbf{Delete}                           & \textbf{Persist}                         & \shortstack{\textbf{Concurrent}                                                                                                         \\\textbf{Operation}} & \textbf{Merge}                               & \textbf{Diff} & \textbf{Aggregation} & \shortstack{\textbf{Concurrent}\\\textbf{Live Branch}} & \shortstack{\textbf{Nested}\\\textbf{Branch}} & \shortstack{\textbf{Schema}\\\textbf{Change}} & \shortstack{\textbf{Physical}\\\textbf{Data Ops}} \\
        \arrayrulecolor{gray!30}
        \midrule
        Neon            & \checkmark                                 & \checkmark                                & \checkmark                               & \checkmark                                   & \texttimes & \texttimes & \texttimes & $\sim$     & \checkmark & \checkmark & \checkmark \\
        Dolt            & \checkmark                                 & \checkmark                                & \checkmark                               & \checkmark                                   & \checkmark & \checkmark & \texttimes & \checkmark & \checkmark & $\sim$     & $\sim$     \\
        TigerData       & \checkmark                                 & \checkmark                                & \checkmark                               & \checkmark                                   & \checkmark & \texttimes & \texttimes & $\sim$     & \checkmark & \checkmark & \checkmark \\
        Xata            & \checkmark                                 & \checkmark                                & \checkmark                               & \checkmark                                   & \checkmark & $\sim$     & \texttimes & $\sim$     & \checkmark & \checkmark & \checkmark \\
        PG file copy    & \checkmark                                 & \checkmark                                & \checkmark                               & $\sim$                                       & \texttimes & \texttimes & \texttimes & \checkmark & $\sim$     & \checkmark & \checkmark \\
        Txn/Savepoint   & \checkmark                                 & \checkmark                                & \texttimes                               & \texttimes                                   & $\sim$     & \texttimes & \texttimes & \checkmark & $\sim$     & $\sim$     & \texttimes \\
    \end{tabular}
    \caption{Feature support matrix for production branchable database systems. \checkmark~indicates full support; \texttimes~indicates no support; $\sim$~indicates partial support.}
    \label{tab:feature-comparison}
\end{table*}

\subsection{Are Nested Transactions Enough?}
\label{sec:taxonomy:txns}

At first glance, existing DBMS features---nested transactions and savepoints---appear sufficient to emulate branching: speculative updates can be executed inside a transaction and either committed or rolled back.
Unfortunately, transaction mechanisms are ill-suited to agentic use cases for many reasons.

First, transactional systems are optimized for short-lived units of work, while long-lived speculative transactions retain historical versions under MVCC and inhibit garbage collection. Although mechanisms such as SAGAs~\cite{garcia1987sagas}, long-lived transactions~\cite{gray1981transaction}, and related  models~\cite{weikum1992concepts,chrysanthis1990acta} decompose long-lived transactions into compensatable steps, they still operate over a single shared ``branch''. In contrast, agentic exploration requires structurally separate branches rather than rollback within a linear execution model.

Second, transactions lack a first-class branch abstraction.
They do not provide durable named branches, explicit lineage relationships, efficient branch switching, nor principled merge semantics.
Merging a selected branch back into a parent state is fundamentally different from committing a transaction when multiple alternative branches have evolved.

%Indeed, a single branch is often approximated by running tentative updates inside a transaction and either committing or rolling back\todo{cite examples in practice (blogs, systems, etc)}.
%\ewu{what does this mean in context of branching? else cut}Some systems also support multiple concurrent transactions at different isolation levels, which can resemble multiple ``worlds''.

Finally, transactions provide logical rollback but do not explicitly treat physical mutations (e.g., creating indexes or materialized views) as speculative state. Agentic exploration often evaluates alternative physical designs, which must persist independently and be compared across branches rather sequentially applied and reverted.

%Second, transactions typically do not provide a durable, named branch abstraction that tracks relationships between branches% lineage () 
%and efficiently switches between branches.
%Finally, transactions lack principled \emph{merge} semantics: combining changes from a selected branch back into a parent state is not equivalent to committing a transaction when multiple alternative branches have evolved.
%While optimistic concurrency control (OCC) also uses a deferred validation phase at commit time, agentic branches 
%Optimistic concurrency control (OCC) does not resolve these issues: while OCC
%avoids lock contention by deferring validation to commit time, agentic branches
%are long-lived and mutate overlapping base state, leading to frequent validation failures and wasted work.
%More fundamentally, OCC remains a transaction mechanism providing no named branch abstraction, no lineage, and no selective merge.

Despite these limitations, it is possible to use transactions to simulate limited exploration use cases that do not need concurrent branches to exist by serially executing the exploration tree depth-first, creating save points at every step, and rolling back.  Our experiments evaluate this as a baseline and find that native database transaction/savepoint serves as a fast light-weight branching mechanism on a limited subset of workflows.
%  will \S\ref{sec:micro} and \S\ref{sec:macro}, we quantify these limitations and show that agentic workloads benefit from dedicated zero-copy branching primitives rather than ad-hoc combinations of transactions and copying.

\subsection{Branchable Database Designs}
\label{sec:taxonomy:zerocopy}

We organize branchable data designs based on the DBMS layer that implements the zero-copy branching mechanism (\Cref{tab:taxonomy}).

\stitle{Storage Substrate: Block-level CoW.}
This approach sits beneath the DBMS and performs copy-on-write over 4-64KB blocks at the storage level. It treats the DBMS as a black-box and is not aware of DBMS-level structures like pages or tuples.

For instance, PostgreSQL 18 introduces a cloning strategy that creates instant database branches by leveraging filesystem-level reflinks\cite{pg-file-copy}. Unlike traditional methods that physically duplicate data, this approach is lightweight because it initially creates a "thin clone" that shares the same physical disk blocks as the source, requiring only minimal DBMS-level configuration to delegate the heavy lifting to the operating system's Copy-on-Write (CoW) mechanisms. However, this efficiency comes with significant constraints: it requires specific filesystem support (like XFS or Btrfs), is limited to cloning databases within the same local PostgreSQL instance, and necessitates terminating all concurrent connections to the source database during the cloning process to ensure a consistent state.

More sophisticated implementations leverage distributed block storage with CoW capabilities. For instance, Tiger's Fluid Storage\cite{tiger-intro} implements a distributed key-value block store where forks track parent blocks via metadata operations so that only modified blocks are copied on write.
Similarly, Xata\cite{xata-intro} uses an NVMe-oF block-storage cluster that splits data into chunks, with each branch maintaining its own metadata index that initially points to shared blocks.
% Vela\footnote{\url{https://vela.simplyblock.io/}} also leverages distributed storage with block-level CoW.

These are lightweight in the data plane but expensive in the control plane as we need to reinstantiate the database compute engine and connections after we create the branch.
Further, coarse-grained CoW (e.g., at storage block boundaries that may not align with database pages) can lead to write amplification when small database updates trigger copies of entire storage blocks.

\stitle{Storage Manager: Page-level CoW.}
This approach targets database pages---such as B$^+$ tree nodes or heap pages---within the DBMS storage manager.
In Minuet~\cite{minuetdb-btree-cow},  branches share a page graph.  When a page is first modified, the storage manager copies it and updates that branch's root pointer. This aligns with DBMS I/O, caching, and buffer pool management.

While this design has potential, these systems are not yet production ready.
Minuet is designed for in-memory B-trees, and exhibits high memory overhead to maintain versions, incurs high garbage collection costs as exploration trees grow, and does not scale to distributed nodes.

%This operates at the natural granularity of database I/O and caching, avoiding block-page mismatch, and integrates with buffer pool management.

%However, we are unaware of production branchable databases implementing page-level CoW. Minuet, designed for in-memory B-trees with writable clones, demonstrated that page-level CoW could enable low-latency transactions and analytics queries to run concurrently on different branches without blocking, supporting efficient "what-if" analysis. However, it faced high memory overhead for maintaining versions, complex garbage collection as lineages grew, and scalability challenges coordinating page-level CoW across distributed nodes.

\stitle{Storage Manager: Content-addressed Tree.}
Systems like Dolt\cite{dolt-intro} store the database using a content-addressed tree (a Prolly tree), where internal nodes and leaves are identified by their hash.  This mirrors git, where a branch is simply a new pointer to a committed root node.  Updates only copy and recompute hashes for the nodes along the path to the updated leaves, and reads traverse the tree starting from the branch's root.

This design efficiently computes diffs and performs three-way merges, but read and write is impacted by large tree fan-outs and chunk sizes; garbage collection must also account for shared chunks across branches.

\stitle{Storage Manager: Delta Overlays.}
Systems like Decibel~\cite{maddox2016decibel} represent a branch as a parent state and a sequence of deltas---represented as a WAL, update log, or set of overlay structures.
Writes append new deltas, while reads apply the deltas to the parent.
Thus, branch creation is cheap and the degree of sharing is proportional to the rate of mutations in the branches.
Decibel explored different physical representations, including version-first (each branch stores deltas in separate segment files forming a chain) and hybrid schemes (combining delta segments with bitmaps for tracking branch membership).

While effective for collaborative data science with long-lived branches, these delta-based approaches face performance challenges under high-frequency branching:
read amplification increases with branch depth, and compaction overhead grows with the number of concurrent branches.

\stitle{Recovery Manager.}
Another approach is to stream the write-ahead log (WAL) to a separate durability service and reconstruct pages for a branch on-demand.
For instance, Neon's\cite{neon-intro} compute nodes stream the WAL to \emph{safekeepers}, i.e., \emph{page servers} that replay WAL segments to construct and serve pages for a given branch.
Branching is implemented by maintaining separate WAL timelines that share a prefix and marking the diverging log entry to avoid copying data.
This effectively provides \emph{page-granularity} branching (via WAL replay) rather than block-level snapshotting.
The design trades write-time copying for read-time reconstruction.
A cold page read on a branch may require replaying a chain of WAL records to materialize the page before it can be served, making performance sensitive to branch depth and whether pages are cached.
In addition, decoupling compute from durability introduces an extra network hop on cold page access, which can increase latency (especially at the tail).
Finally, because correctness requires retaining historical WAL (and associated page versions) until all dependent branches no longer need it, long-lived branches can delay garbage collection and lead to storage bloat.

%exposes a page-oriented sharing boundary (shared history + CoW in WAL) rather than block-level snapshots.

%represent a branch as a pointer 
%A branch can be represented as a pointer to a root node in a content-addressed
%tree (e.g., a prolly tree or Merkle DAG), where internal nodes and leaf chunks
%are identified by their hash.
%Creating a branch is a metadata operation, a new root pointer into the same
%chunk graph, and writes produce new chunks only for the modified paths,
%automatically sharing all unmodified subtrees with the parent.
%Reads traverse the tree from the branch's root without applying overlays or
%consulting parent state.
%
%Dolt is a concrete instantiation of this approach, implementing a SQL database
%over a prolly-tree storage engine with Git-style commits and branches.
%

\stitle{Transaction and Logical Layers.}
In principle, higher DBMS layers can support branching as well.  For instance, MVCC already maintains tuple version chains for isolation, and may be extended with branch identifiers and by treating each branch as a long-lived snapshot.  However, to the best of our knowledege, this design is not implemented.   Similarly, ServiceWeaver~\cite{Zhu2025ViveLD} models each branch as a set of views that store inserts and deletes for each relation, and rewrites queries using triggers.  However, it is primarily designed for logical differential testing for two versions of a codebase and not agentic exploration use cases that involve arbitrary branch trees that contain physical and logical mutations.

\subsection{Feature Comparison}

Table~\ref{tab:feature-comparison} summarizes the branching capabilities of production systems. We exclude research systems (Minuet, Decibel, ServiceWeaver) from this comparison as they are not production-ready and lack comprehensive feature support or availability for empirical evaluation. The table reveals that while all production systems support basic branch creation, deletion, and persistence, significant gaps remain: no system supports cross-branch aggregation (querying data across multiple branches simultaneously), and only Dolt provides first-class merge and diff operations due to its Git-inspired content-addressed design.

% \stitle{Transaction Manager.}
% MVCC is the most widely implemented concurrency control mechanism, and already maintains multiple versions of tuples to support isolation.
% Branching can be achieved by extending MVCC with a \emph{branch identifier} and treating each branch as a long-lived snapshot that can accept new versions.

% \stitle{Logical Views.}
% Systems like \ewu{NAMES} simulate branches by logically creating derived view(s) that encode the mutations.  For instance, \ewu{EXAMPLE}.
% While this does not require DBMS modifications, this fails to handle physical operations (e.g., index creation), do not support arbitrary updates, and often cannot capture the \ewu{what does this mean?}execution-time behavior that tuning and scenario workloads require.

\section{\textbf{The \sys\space Benchmark}}
\label{sec:benchmark}

We now describe the design of \sys, which exercises the requirements in the previous section and can be parameterized to emulate our motivating scenarios.
We describe the database schema, execution model, and parameterizations for each scenario.

%We design a parametrized macrobenchmark that captures the end-to-end agent
%workflow. By varying the worker count, branching topology, mutation mix, and background
%maintenance parameters, the benchmark template can be instantiated to emulate
%each of the above five workflows. We define a shared base schema, an
%automata-based execution model, and per-workflow instantiations with concrete
%operations.

\begin{figure}[h!]
    \centering
    \includegraphics[width=\columnwidth]{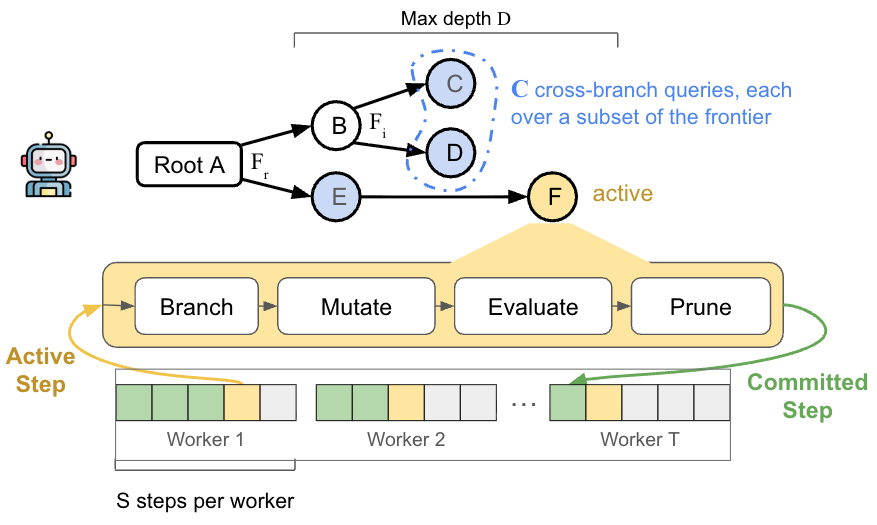}
    \caption{Architecture of the \sys ~~ macrobenchmark showing the branch-mutate-evaluate-prune loop executed by parallel workers.}
    \label{fig:macro-flow}
\end{figure}

\subsection{Database Setup.}
We adopt the CH-benCHmark schema~\cite{cole2011mixed}, which unifies the
TPC-C and TPC-H benchmarks into a single hybrid workload.
The schema retains TPC-C's nine transactional tables (e.g., \texttt{warehouse},
\texttt{customer}, \texttt{order}, \texttt{orderline}, \texttt{stock}) and
augments them with a subset of TPC-H's analytical tables (i.e., \texttt{region},
\texttt{supplier}, and a slightly modified \texttt{nation}) to support complex
analytical queries over the transactional data.
% The schema is scaled by warehouse count $W$: each warehouse serves 10 districts
% with 30K customers each, and \texttt{stock} maps all 100K items to each
% warehouse, yielding $100\text{K} \times W$ stock rows.
This combination of transactional tables with rich foreign-key relationships
and large tables for analytics forms the basis for the scenarios: vibe coding features for an inventory application, failure reproduction over the transaction sequences, data curation, fulfillment that optimizes for shipping costs, and simulating inventory shortfalls.
%analytical query patterns naturally exercises the full range of mutations
%and evaluations across our five workflows: schema migrations and integrity
%tests, replaying and mutating transaction sequences, bulk-derived table
%rewrites with full-scan validation, lightweight counterfactual updates
%with analytical reward queries, and forward simulation of order fulfillment
%scenarios over the inventory and order tables.

\subsection{Execution Model.}

\Cref{fig:macro-flow} illustrates the architecture of the benchmark, with concrete operations  executed in each phase.
$T$ concurrently executing worker threads are each assigned $S$ steps of work.  
In the figure, the green steps denote completed steps, yellow denotes the currently executing step, and gray denotes future steps.

The workers all execute on a shared branch tree that starts from the root schema state ($A$ in the figure).   The tree's shape is defined by the root fanout  $F_{\text{r}}$, which specifies the number of children the root node should have;  the inner fanout $F_{\text{i}}$, which bounds the number of children interior nodes have; and a maximum depth~$D$ below root.
These parameters allows the benchmark to express a range of tree topologies.  For instance, $D{=}1$ forces a flat ``star'' shape, $F_{\text{i}}{=}1$ forces chains below the root, and intermediate values yield bounded bushy trees.

Each step corresponds to a sequence of parameterized operations that operates in four phases:
\begin{enumerate}[leftmargin=*]
    \item \textbf{Branch}: selects a node whose child count is below
          its fanout limit and whose depth (below root) is below~$D$, 
          forks a new branch, and connects to it.
    \item \textbf{Mutate}: apply $M_{\text{s}}$ DDL operations (e.g., schema changes) and $M_{\text{d}}$ data mutations on the new branch.
    \item \textbf{Evaluate}: execute $Q_{v}$ read queries on the branch to
          validate or score the result.
    \item \textbf{Prune}: with probability $\gamma$, delete the branch.
\end{enumerate}
\noindent \Cref{a:macroqs} lists examples of mutation and evaluation statements for each workflow.  A branch is considered \emph{committed} once it has completed all phases of a step and has not been pruned.    We call all of the committed leaves the {\bf Frontier} of the tree (denoted as \textcolor{blue}{blue} nodes in the figure.

In addition to per-step work, a workflow executes $C$ cross-branch queries that are executed evenly throughout the workflow execution; these are not considered as one of a worker's $S$ steps.  When a cross-branch query is ready to run, it is selected by the first active thread on a FIFO basis.  % spread evenly across the $S$ steps  and performed by threads.
Each cross-branch query reads across a subset of the frontier (committeed leaves).   When $C{=}1$, a single cross-branch aggregation runs at the end of the workflow when all workers have completed their $S$ steps.
After all threads finish their steps and the final cross-branch queries
complete, the workflow is complete.

\stitle{Core Parameters.} We organize the benchmark parameters into two groups.   \Cref{tab:workload-params} lists the configuration for each workflow.
\begin{itemize}[leftmargin=*]
    \item \emph{Setup:}
          Workers $T$, steps per worker $S$,
          cross-branch queries $C$ (spread evenly across all $S$ steps;
          each reads across committed leaf branches),
          root fanout $F_{\text{r}}$, inner fanout $F_{\text{i}}$, and
          maximum depth $D$ below the root level.
          Separating $F_{\text{r}}$ from~$F_{\text{i}}$ captures workflows
          whose root branches at a different rate than deeper levels.
    \item \emph{Per-step:}
          Schema changes $M_{\text{s}}$,
          data mutations $M_{\text{d}}$,
          evaluation queries $Q_{v}$, and
          prune probability $\gamma$.
\end{itemize}

\subsection{Branch Lifecycle Microbenchmarks}
We also develop a microbenchmark suite to isolate the overhead of \emph{branch lifecycle primitives} from the cost of data- and schema-level SQL work.  For instance, how are reads impacted by the number of existing branches?    What is the cost of creating and connecting to a new branch?
While macrobenchmarks measure end-to-end workflow performance, microbenchmarks provide controlled, fine-grained measurements of individual branch operations (e.g., create, switch, delete) and their interaction with SQL execution.  

To make it easy to add and evaluate new backends, we abstract branchable DBMS backends behind four high-level operations that, respectively, create a new branch and return an identifier, connect to a named branch, delete a named branch, and execute SQL statements on the active branch:
\begin{verbatim}
  id = create_branch(); 
  connect_branch(id); 
  delete_branch(id);
  execute_sql(qs);
\end{verbatim}

\noindent The microbenchmark driver executes a scripted sequence of these calls and times each operation independently.
Each microbenchmark is specified by a configuration file that defines (i) a \emph{setup phase} (untimed) used to construct the initial state (e.g., base data size, initial branching structure), and (ii) an \emph{execution phase} (timed) that issues the operations to be measured.
This separation ensures that reported results reflect steady-state primitive costs rather than one-time initialization overheads.

\begin{table*}[!ht]
    \centering
    \small
    \begin{tabular}{l l r r r r r}
       \arrayrulecolor{gray!50}
        \textbf{Group} & \textbf{Parameter}            & \textbf{Software\ Dev} & \textbf{Failure\ Repro} & \textbf{Data Cleaning} & \textbf{MCTS} & \textbf{MC Simulation.} \\
        \midrule
        \multirow{7}{*}{Setup}
                       & Workers $T$                   & 5                      & 1                       & 10                     & 10            & 1000                    \\
                       & Steps/Worker $S$               & 20                     & 10                      & 20                     & 100           & 1                       \\
                       & Cross-branch queries $C$      & 1                      & ---                     & 2                      & ---           & 1                       \\
                       & Root fanout $F_{\text{r}}$    & 5                      & 10                      & 10                     & 10            & 1000                    \\
                       & Inner fanout $F_{\text{i}}$   & 3                      & ---                     & 3                      & 10            & ---                     \\
                       & Depth $D$                     & 3                      & 1                       & 3                      & 25            & 1                       \\
        \midrule
        \multirow{4}{*}{Per-step}
                       & Schema changes $M_{\text{s}}$ & 1                      & 5                       & 1                      & ---           & ---                     \\
                       & Data mutations $M_{\text{d}}$ & 1                      & 45                      & 1                      & 1             & 50                      \\
                       & Read queries $Q_{v}$          & 2                      & 1                       & 1                      & 1             & 1                       \\
                       & Prune prob $\gamma$           & 0.1                    & 1                       & ---                    & 0.1           & 1                       \\
    \end{tabular}
    \caption{Parameter instantiations for each workflow (--- means no such operation).}
    \label{tab:workload-params}
\end{table*}

\subsection{Evaluation metrics.}
Our metrics seek to quantify end-to-end branchable database performance across workflows and to isolate performance bottlenecks for primitive branch operations.

\subsubsection*{End-to-end workflow time} Total wall-clock time to
          complete a full workflow all $S$ steps from $T$ threads plus
          the cross-branch aggregation.
          This integrates branching overhead,
          mutation throughput, schema change performance, and reclamation into a
          single number, enabling direct comparison across backends for the
          same workflow configuration.
\subsubsection*{Branching overhead ratio} Fraction of total workflow
          wall-clock time spent on branch management (creation, deletion,
          switching) versus productive work (mutations and evaluation
          queries). A high ratio indicates that branching infrastructure, rather than
          the workload itself, is the bottleneck.
          Comparing this across backends and workflow types reveals
          which systems amortize branching costs effectively.
\subsubsection*{Storage Efficiency}
We consider both (i) how much additional storage branching actually consumes and (ii) how quickly disk space is reclaimed after branch deletion.
Workflows that repeatedly create and discard branches are especially sensitive to both effects.
If branches induce substantial storage amplification, storage can balloon even when the live branch count is small; and if reclamation is slow, storage can grow monotonically over time even when the logical branch count remains bounded.
\subsubsection*{Per-operation latency} Measures operation latency at the granularity of individual branching primitives (branch lifecycle operations), data operations (DML), and schema operations (DDL) to attribute end-to-end workflow performance to specific system mechanisms.

    % \item \textbf{Storage amplification.} Peak disk consumption divided by
    %       the base dataset size.
    %       An efficient system scales storage proportionally to the
    %       actual mutation delta across branches; an inefficient one copies
    %       entire tables regardless of how much data actually changed.
    % \item \textbf{Isolation overhead.} Evaluation query latency on a single
    %       branch as the total number of \emph{other} active branches
    %       increases.
    %       Measures whether the system maintains performance isolation or
    %       whether unrelated branches degrade read and write throughput
    %       through shared resources (buffer pool, catalog locks, storage I/O).

% \end{itemize}

% \input{sections/macro.tex}
\section{Evaluation}
\label{sec:evaluation}

We now evaluate our macro and microbenchmarks on representatives of modern branchable database systems.   We first evaluate branching costs for each system, and then evaluate the extent that each system is capable of running the macrobenchmarks.

We find that not all systems are capable to run the operations that the macrobenchmark workflows require---likely because branching is a relatively new capability---and none can run the workflows to completion within 2 hours.  For this reason, we then restrict the evaluation to the two dominating systems---Neon and Dolt---and use a scaled down macrobenchmark to study their operation costs.  We find that their main query performance difference lies in range queries and DDLs, so we then report microbenchmark results on these operations.   Overall, we find that modern branchable databases make architectural trade-offs to support very distinct workflow classes.

\subsection{Experiment Setup}
% kk: cutting the following as we already say that our schemas are based on CH-bencCHmark
%We use the CH-benCHmark schema~\cite{cole2011mixed}, a hybrid OLTP/OLAP benchmark combining TPC-C transactional tables with TPC-H analytical queries.
%We generate the databases using scripts provided by Ch-benCHmark, converting the CSVs into SQL dumps so that the databases are consistently imported into each DBMS.

\subsubsection{DBMS Systems}
We evaluated on Neon, Xata, TigerData, Dolt, and two PostgreSQL approaches (file copy and transactions/savepoints).  Since many of the systems are relatively new, we now describe some of the engineering challenges that arose during benchmark evaluation.

For PostgreSQL transaction/savepoint and PostgreSQL file-copy clone, we run Postgres 18 with \texttt{max\_connections} set to 1200.

Neon is both a cloud-hosted service as well as an open-source project.  However, we were unable to properly run Neon locally despite considerable engineering effort. Their results are with respect to their hosted service, which we ensure runs in the same region as our VM (AWS us-east-1).

Dolt is self-hosted on AWS, which we built from source. We ran into psycopg2 errors during the initial experiment. After reporting those to the Dolt team, they confirmed and fixed the bug~\cite{dolt-bug}.

% \ewu{All the pain associated with getting each service to work, to collect metrics, and so on.  All of these are important as part of delivering a useful system.}

All self-hosted experiments (Dolt, PostgreSQL) run on a single \texttt{c8i.4xlarge} EC2 instance (16 vCPUs, 32 GB RAM) with an 150 GB EBS volume in \texttt{us-east-1}.
All hosted services (Neon, Tiger, Xata) implement  rate-limiting to various degrees. We implemented basic timeout and retry mechanisms in order to avoid immediate failure when rate limited.

\subsubsection{Macrobenchmark Setup}
We find that most systems are not able to complete the full benchmark and thus we developed two variations:
\begin{enumerate}[leftmargin=*]
    \item \textbf{Full.} Our target configuration with scale factor $W=5$ (five warehouses) and the full workflow parameters from Table~\ref{tab:workload-params}. This setting measures how well each system scales under realistic agentic exploration, surfacing performance bottlenecks and resource limitations.
    \item \textbf{Mini.} A reduced configuration with scale factor $W=1$ (one warehouse) and downscaled workflow parameters (fewer workers and fewer steps), while keeping the per-step operation mix unchanged.
          This setting primarily tests whether each DBMS can execute the required branch and SQL operations for each workflow and exposes correctness/feature gaps and per-operation costs.
\end{enumerate}

\subsubsection{Microbenchmark Setup}
The microbenchmark always runs with scale factor $W=1$.  We first study the cost of branching across all systems. From our macrobenchmark experiments, we found that the two ``production-ready'' DBMSes, Neon and Dolt, are comparable for almost all data operations with the exception of DDLs and range queries, so we focus on reporting their performance.

\subsection{Branching Performance}

\begin{figure}
    \centering
    \includegraphics[width=\columnwidth]{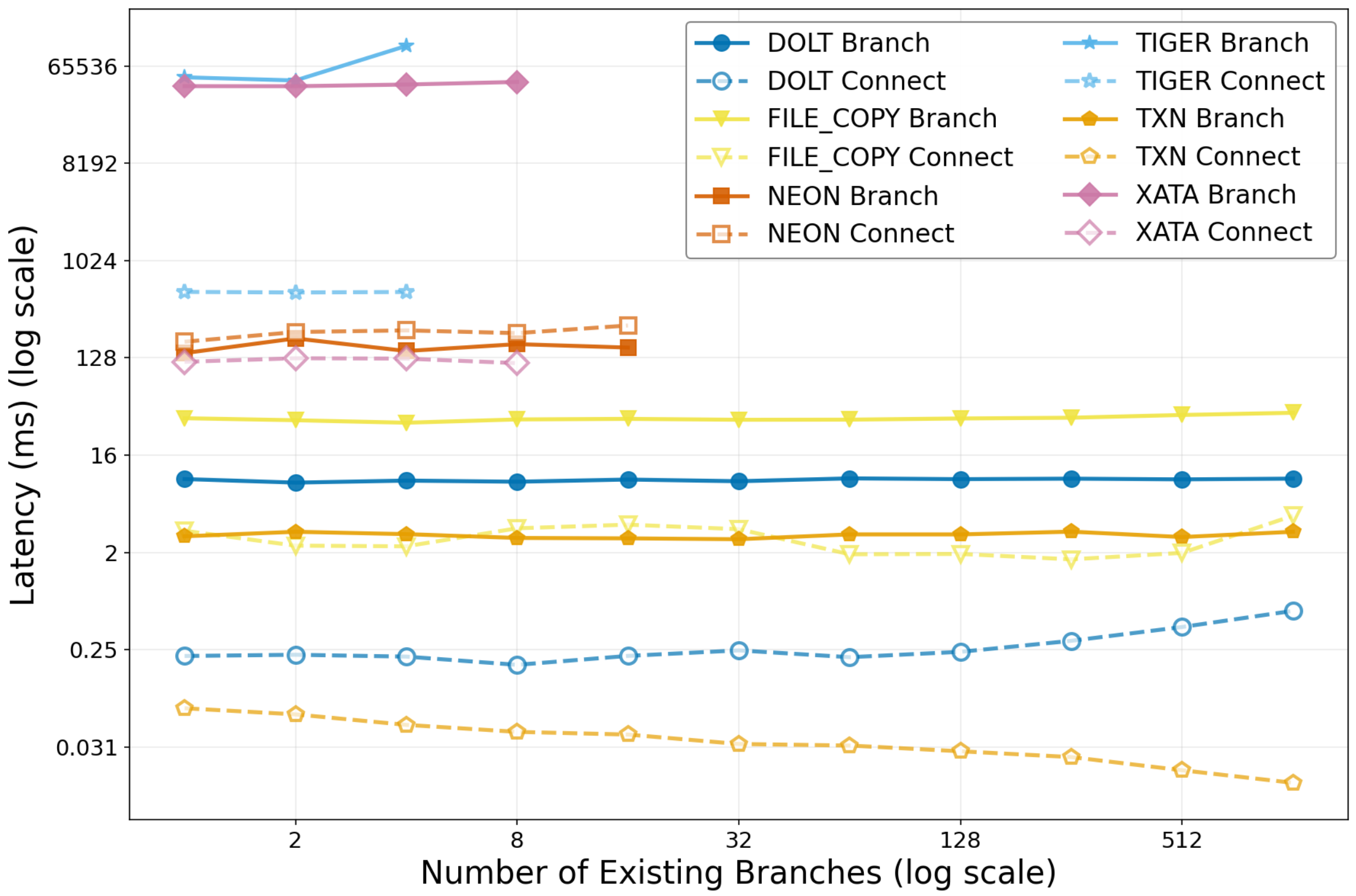}
    \caption{Branch creation latency on n-th branch, across all systems.}
    \label{fig:latency-branch-all}
\end{figure}

Every agentic application requires the ability to create and connect to new branches.
Thus, before running any agentic workloads, we first measure branch and connect performance for all systems.

\subsubsection{Single-Thread}
We repeatedly create and connect to a new branch, and create a spine shaped branch tree (fanout=1) where each new child branch uses the previous branch as parent.
\Cref{fig:latency-branch-all} reports the latency to create and connect to the n-th branch on a single thread.  Xata and Tiger have orders of magnitude higher branch creation latencies compared to all other systems---to the extent that we disqualify these systems for the rest of the experiments.  Neon is much slower than Dolt, mostly because Neon's architecture provisions a new compute instance for every branch to act as the branch's page server and partly due to network delays to Neon's cloud service.
As expected, PostgreSQL transaction creation forms the lower bound.

\begin{figure}
    \centering
    \includegraphics[width=\columnwidth]{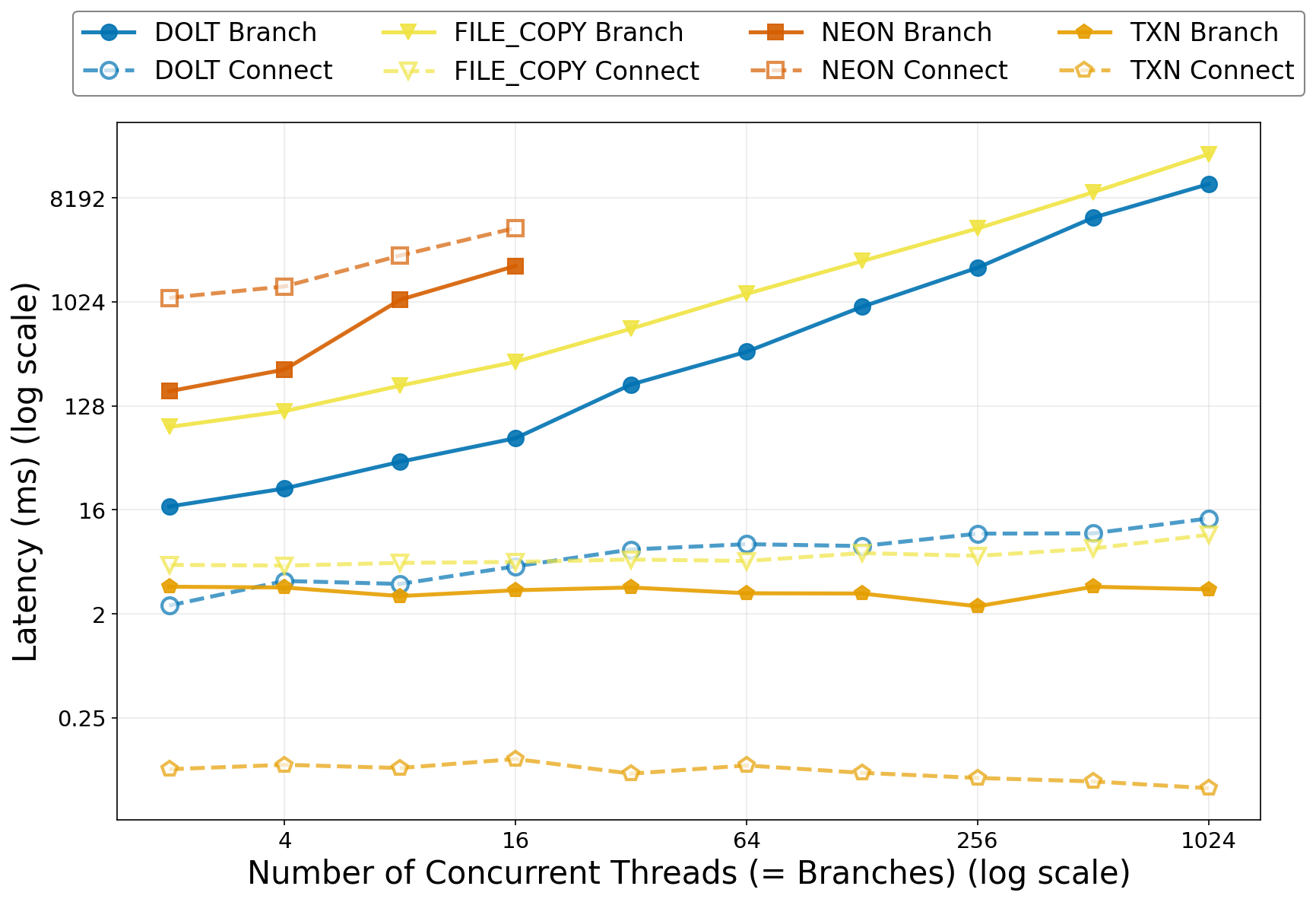}
    \caption{Multi-branch connection performance across systems.}
    \label{fig:multi-branch-connect}
\end{figure}

\subsubsection{Multi-Thread}
To measure interference, we evaluate the performance of increasing the number of threads that concurrently create and connect to new branches.  Each thread branches from the root to form a wide, depth=1 tree.
As show in \Cref{fig:multi-branch-connect},
%We create a fan\_out (star) shaped branch tree where each child branch is created from a common parent.
%The branches are created in parallel, and each thread works on creating and connecting to a different branch.
with the exception of PostgreSQL transactions, the cost of creating {\it and connecting} to new branches grows with the number of concurrent branches in all systems.

\subsection{Backend Capability Analysis}
\label{sec:capability}

\begin{table}
    \centering
    \footnotesize
    \begin{tabular}{l c c c c c}
        \arrayrulecolor{gray!30}
        \textbf{System} & \shortstack{\textbf{Software}                                                     \\\textbf{Dev}} & \shortstack{\textbf{Failure}\\\textbf{Repro}} & \shortstack{\textbf{Data}\\\textbf{Cleaning}} & \textbf{MCTS} & \textbf{Simulation} \\
        \midrule
        Neon            & $\sim$                        & \blue{\checkmark} & $\sim$     & $\sim$     & \red{\texttimes}     \\
        Dolt            & \blue{\checkmark}                    & \blue{\checkmark} & \blue{\checkmark} & $\sim$     & \red{\texttimes}     \\
        Xata            & \red{\texttimes}                    & \red{\texttimes} & \red{\texttimes} & \red{\texttimes}          & \red{\texttimes}          \\
        TigerData       & $\sim$                             & \blue{\checkmark}          & $\sim$          & $\sim$          & \red{\texttimes}          \\
        \midrule
        PG Clone        & \red{\texttimes}                    & \blue{\checkmark} & \red{\texttimes} & \red{\texttimes} & \blue{\checkmark} \\
        Txn/Savepoint   & \red{\texttimes}                    & \blue{\checkmark} & \red{\texttimes} & \red{\texttimes} & \blue{\checkmark} \\
    \end{tabular}
    \caption{System capability matrix showing which systems can execute the workflow on full configuration. \blue{\checkmark}~indicates capable; \red{\texttimes}~indicates incapable; $\sim$~indicates partially capable with errors or crashes.}
    \label{tab:system-capability}
\end{table}

We use the macrobenchmark to evaluate the features that each system supports, and limited each workload's execution to 2 hours.
Table~\ref{tab:system-capability} shows that no system was able to fully complete the five agentic applications within the 2 hours.
The root causes of failed executions are based on whether a given system supports features such as mutations in concurrent branches, persistent named branches, efficient data operations, or simply stable API support. This subsection analyzes the full and partial execution failures.

\subsubsection{Full Failures}
We denote a \red{\texttimes} in the table when a system was unable to execute
a workflow.

\stitle{PostgreSQL} transactions are sufficient only for the Failure Reproduction and Simulation workflows where a branch is immediately pruned after evaluation ($\gamma=1$).
This is because switching to a different savepoint or transaction context requires reverting all work done since that savepoint, effectively discarding the current state.
Similarly, PG Clone's limitation is that the source/template database cannot have any active connections at the time of cloning, because PostgreSQL blocks all connections to the template database during the clone operation to ensure a consistent, crash-free copy.
Active connections cause cloning to block.

\stitle{Xata} in principle should be able to support the Failure Reproduction workflow.  However, during the course of experiments, their team disabled our API access, which subsequently reported 403 failures.

\stitle{Neon} finished 348 out of 1000 steps for the Simulation workflow, however we consider it a full failure because not all scenarios were completed and the final outcome requires a cross-branch aggregation across {\it all} simulated scenarios.

\stitle{Dolt's} reads must traverse its content-addressed tree, which grows longer with the number of branches and mutations.    This means that as the number of concurrent branches grows, the read performance becomes considerably slower due to the number of random seeks.    For simulation, which runs upwards of 1000 concurrent branches, the reads become so slow that Dolt did not terminate gracefully at the time out.  For these reasons, we were not able to collect final statistics about the state of the database for this workflow.

\begin{figure}
    \centering
    \includegraphics[width=\columnwidth]{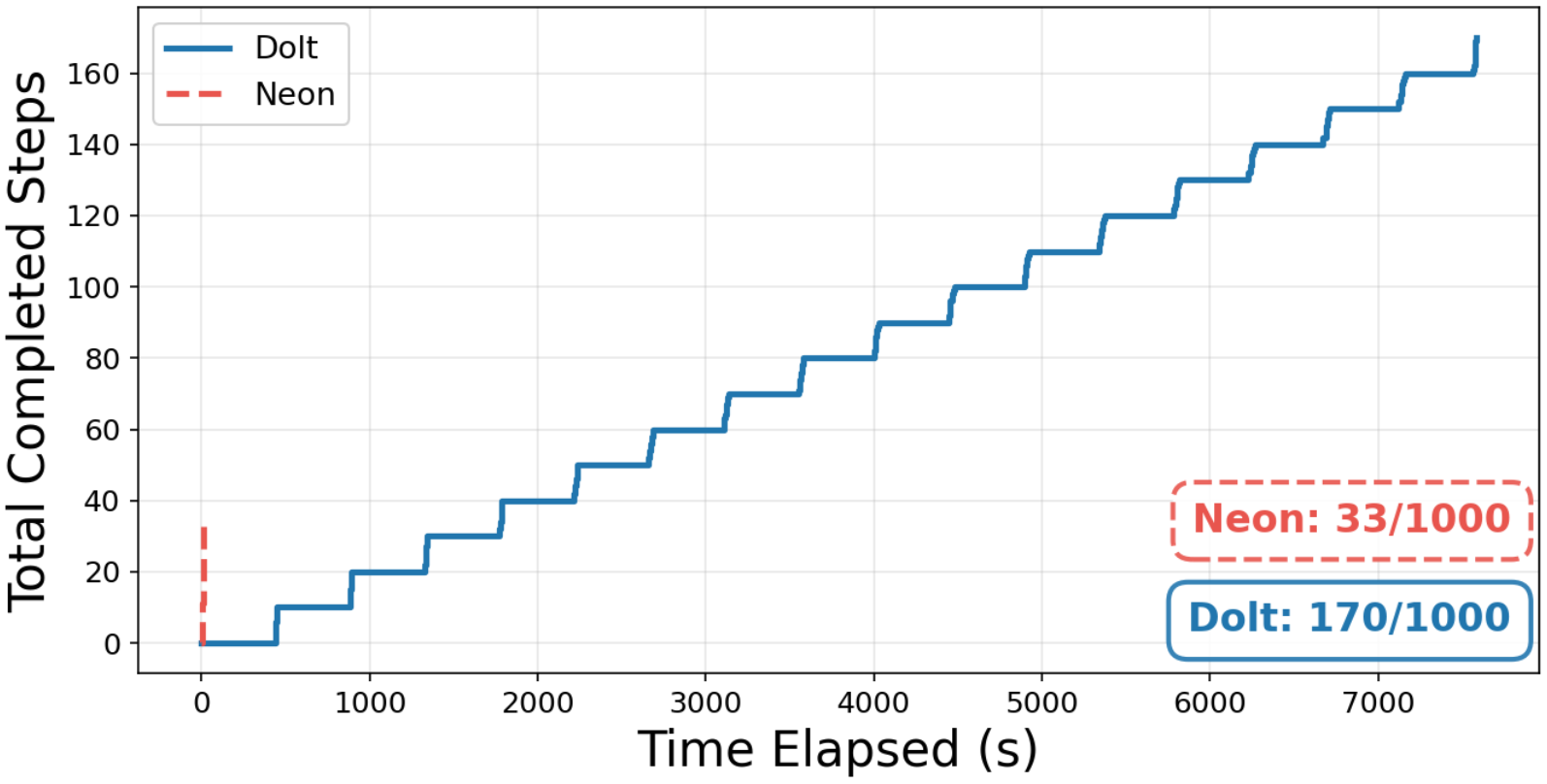}
    \caption{Unfinished MCTS exploration tree performance.}
    \label{fig:unfinished-mcts}
\end{figure}

\subsubsection{Partial Workflow Execution}

We use $\sim$ to denote a system that completes a subset of the workload but either did not finish or encountered errors during the execution.

\stitle{TigerData} did not complete any workflows other than Failure Repro. Data Cleaning completed 16/200 steps but timed out waiting for additional branches.   The remaining workflows failed due to undocumented API rate-limiting causing API calls to fail. TigerData's complete isolation between services ensures consistently fast on-branch CRUD operations, but the bottleneck imposed by branch creation time effectively throttles performance in our benchmark suite enough to prevent complete experiment inclusion.

\stitle{Neon's} cloud service does not support more than 20 concurrent ``live'' branches because each branch is attached to a compute instance.   Although this ensures that queries to each live branch is fast, many workloads require more than 20 concurrent branches: Software Dev, Data Cleaning, and MCTS.
When the benchmark tries to acquire a new branch and the API returns an error, we mark that step as a failure and move on to another step.

%Dolt encountered partial failures for MCTS and Simulation because their read operations became very expensive.  As the number of branches and data mutations increase, Dolt's reads require longer traversals across its content-addressed tree and caused workloads to time out.   For simulation, which runs upwards of 1000 concurrent branches, the reads become so slow that Dolt did not terminate gracefully at the time out.
\stitle{Dolt's} partial failure for MCTS is because its reads become increasingly expensive.  Unlike Simulation, MCTS runs far fewer concurrent branches and Dolt terminated gracefully at 2 hrs.

We use MCTS as an illustrative example of their execution progress over time.  Figure~\ref{fig:unfinished-mcts} shows the number of completed steps (y-axis) over time (x-axis) up until the 2 hour timeout.    We see that their progress is step-wise linear.   Although Neon completes the first several steps quickly, it then blocks due to the concurrent live branch limit.
In summary, Neon only completed 25/100 steps for Software Dev, 26/200 for Data Cleaning, 33/1000 for MCTS, and 348/1000 for Simulation; Dolt completed 170/1000 for MCTS and an unknown number for simulation (see above).

\subsubsection{Storage Utilization}
We attempted to measure storage utilization statistics for each system at the end of a workflow's execution or time out.   Since Xdata, TigerData, PG Clone, and PostgreSQL Transactions fully failed on almost all workloads, we focus on Neon and Dolt.
\Cref{tab:storage_delta} reports storage overhead that we measured at the end of executing each application (or upon its timeout).

\begin{table}
    \centering
    %\footnotesize
    \begin{tabular}{r r r r r r}
        \arrayrulecolor{gray!30}
        \textbf{System} & \shortstack{\textbf{Software}                                    \\\textbf{Dev}} & \shortstack{\textbf{Failure}\\\textbf{Repro}} & \shortstack{\textbf{Data}\\\textbf{Cleaning}} & \textbf{MCTS} & \textbf{Simu.} \\
        \midrule
        Dolt            & 93.1 MB                       & 63.6 MB & 94.1 MB & 3.8 MB & --- \\
        Neon            & 4.0 GB                        & 0 B     & 4.7 MB  & 4.1 MB & --- \\
    \end{tabular}
    \caption{Storage overhead per workflow for Dolt and Neon.}
    \label{tab:storage_delta}
\end{table}

\stitle{Neon} spends over $43\times$ more storage overhead for the Software Dev workload because the application executes schema updates with data backfills.   This suggests that Neon is not optimized to efficiently manage data overlaps upon schema updates.

\stitle{Dolt} consumes considerably more storage overhead in Failure Repro and Data Cleaning.   The former is because Dolt stores the full history of its data even for pruned branches, whereas Neon reclaims the resources.   The latter is because the workload adds columns with \code{DEFAULT FALSE}; Neon's PostgreSQL engine stores the \code{FALSE} as metadata, whereas Dolt materializes the column.

\stitle{Caveat.}  We emphasize that the storage statistics for {\bf Neon are not fully reliable}, and so they should be interpreted for order-of-magnitude comparative purposes only.  This is for two reasons.  The first is that Neon and Dolt did not finish the same number of steps in each workflow, so the exact numbers are not apples-to-apples.  The second, more importantly, is that Neon's metrics reporting has limitations.   After starting an experiment run, Neon only begins collecting metrics 15 minutes into the experiment, and then updates the statistics at the top of each hour (e.g., 1PM, 2PM).  For this reason, we are unable to analyze storage overheads during a benchmark, and due to resource reclamation, may under-report storage overheads for workflows that prune branches. In contrast, we ran Dolt locally and have precise statistics.

\begin{figure}[h]
    \centering
    \includegraphics[width=\columnwidth]{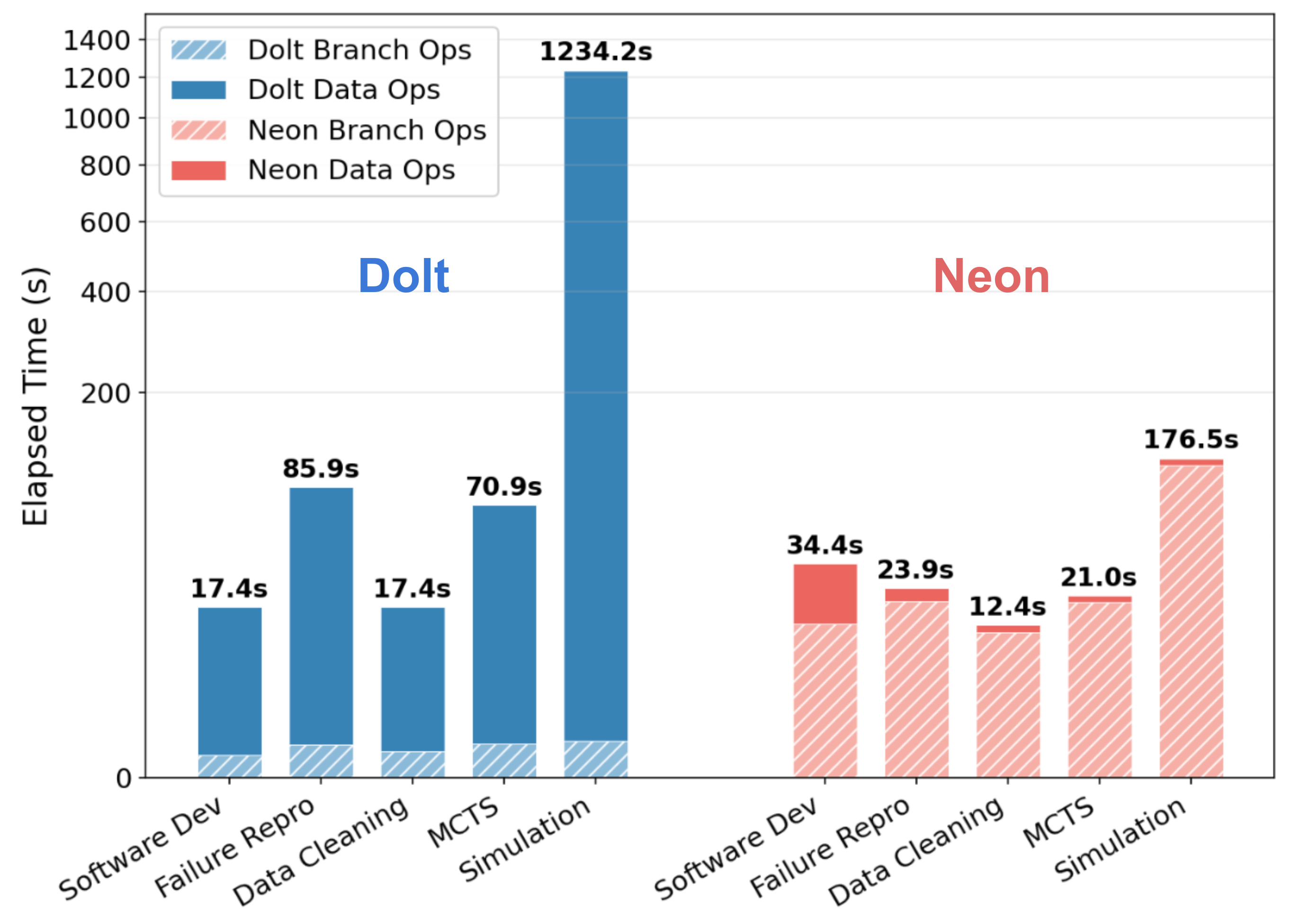}
    \caption{End-to-end latency for each workflow (mini config).}
    \label{fig:e2e-latency}
\end{figure}

\begin{figure*}
    \centering
    \includegraphics[width=\textwidth]{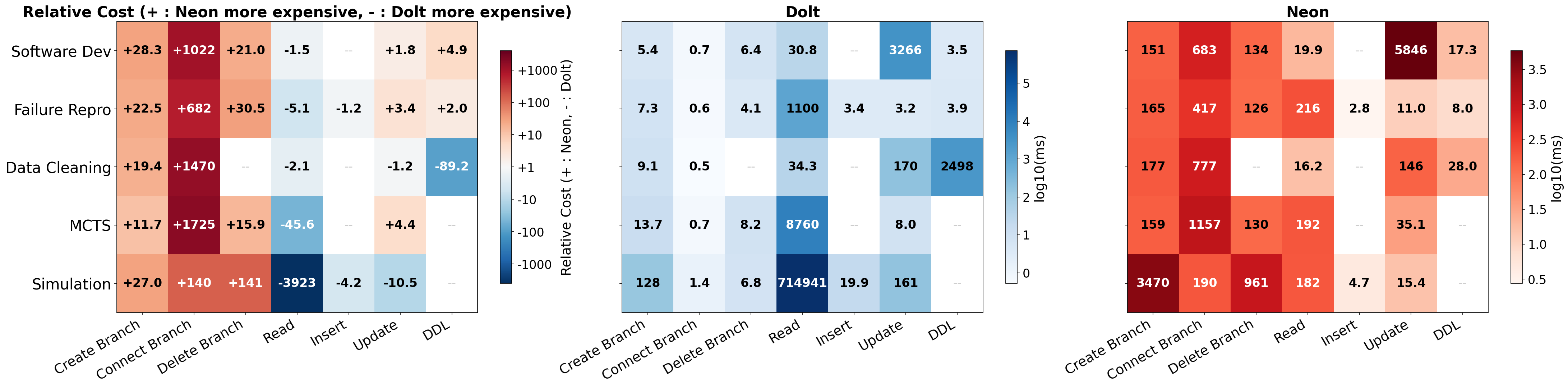}
    \caption{Median latency heatmap across systems and operations.}
    \label{fig:median-heatmap}
\end{figure*}

\subsection{Backend Performance Analysis}
Based on the capability analysis in \S\ref{sec:capability}, we focus our performance evaluation on Neon and Dolt, the two systems that could execute the broadest set of workflows (Table~\ref{tab:system-capability}) while being reasonbly fast at branching. We use the mini configuration to ensure both systems
complete all workflows within the timeout, enabling controlled comparison without API-imposed limits or multi-hour query execution.

\subsubsection{Latency Breakdown by Workflow}
Figure~\ref{fig:e2e-latency} shows end-to-end workflow latency decomposed into data operations (SELECT, INSERT, UPDATE, DDL) and branch lifecycle primitives (create, connect, delete).

Dolt's branch operations are very fast, but it is limited by slow reads due to the need to traverse the content-addressed tree's structures---which grows with the number of branches and mutations---for each query. We see this most dramatically with Simulation, where the high number of branches (1000) and mutation operations (50 per step) causes reads to be particularly expensive.      However, Dolt is still faster when the number of mutations is low (Software Dev).   Neon's branching operations are dominated from network round-trips to the control plane API---even when run in the same AWS region---and per-branch compute provisioning delays.  The data operations are fast because each branch is provisioned with a dedicated compute instance.

% We won't able to get a storage profile for Neon in the mini config due to its
% cloud storage reporting window issue (1hr window only and the workflows finish in
% way less time).

\subsubsection{Heatmap for Neon and Dolt}
Figure~\ref{fig:median-heatmap} compares median latency for each type of operation (x-axis) and each workflow (y-axis).    The left-most heatmap reports the relative performance between Neon and Dolt, where positive cells are where Neon is slower and negative is where Dolt is slower.  The middle and right plots report absolute latencies in milliseconds

We find that most data operations are relatively similar, except for DDL statements and reads.   Dolt's DDL operations in the Data Cleaning workflow incur substantially higher cost than Neon because Dolt materializes DEFAULT column values for all existing tuples, while PostgreSQL treats this as a metadata-only operation.

Dolt's read latency is $2-4000\times$ higher than Neon's across workflows---particularly for range read queries.  This motivates our microbenchmarks to focus on read performance to understand the performance under varying branch and concurrent thread counts.

\subsection{Microbenchmark: Drill down on reads}

\begin{figure}[h]
    \centering
    \includegraphics[width=\columnwidth]{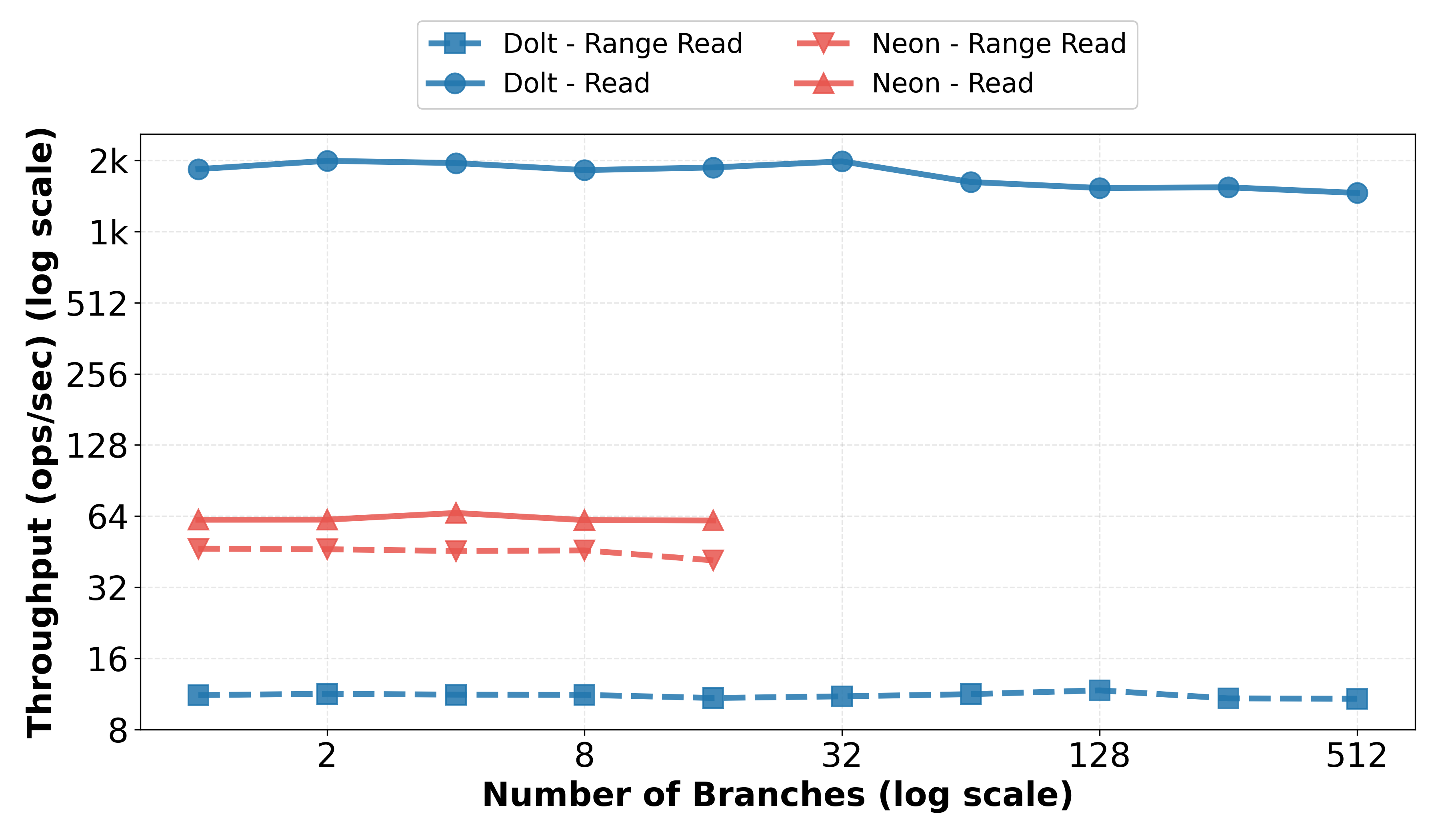}
    \caption{Throughput for a single thread as number of branches increase (range size = 100).}
    \label{fig:tp-br}
\end{figure}

To understand the architectural trade-offs between Neon and Dolt revealed by the macrobenchmark, we evaluate read performance scaling in a microbenchmark along two dimensions: query selectivity (point vs range) and concurrency (single vs multi-thread).
We use the CH-benCHmark schema (W=1), where point reads query a single key and range reads query a contiguous 100-key range.
We vary the number of branches (x-axis) and measure throughput in operations per second (y-axis) with synchronous request execution, consistent with the macrobenchmark pattern.
In the single-thread experiment, one thread operates on a single branch to isolate the effect of branch count on per-branch performance.
In the multi-thread experiment, each branch is accessed by one thread to measure aggregate system capacity as concurrent branches scale.

\stitle{Query Selectivity: Point vs Range.}
Dolt outperforms Neon on point reads across all configurations because Dolt's point reads execute locally, whereas Neon incurs network round-trip overhead to the cloud service.
However, Dolt's range scan performance degrades significantly---$2-4000\times$ slower than Neon---due to the cost of traversing its content-addressed tree structure.
% Each range query requires multiple random seeks through the tree, and this traversal cost compounds with the number of branches and mutations in the system.
This explains Dolt's poor performance on Simulation (1000 branches with 50 mutations per step) where range queries dominated end-to-end latency.

\stitle{Single-Branch Scaling.}
Figure~\ref{fig:tp-br} shows that increasing the number of branches in the system does not degrade single-branch read performance for either system.
This indicates that both architectures maintain branch isolation: Neon's per-branch compute instances operate independently, and Dolt's content-addressed tree does not incur cross-branch interference when only a single branch is active.

\begin{figure}[h]
    \centering
    \includegraphics[width=\columnwidth]{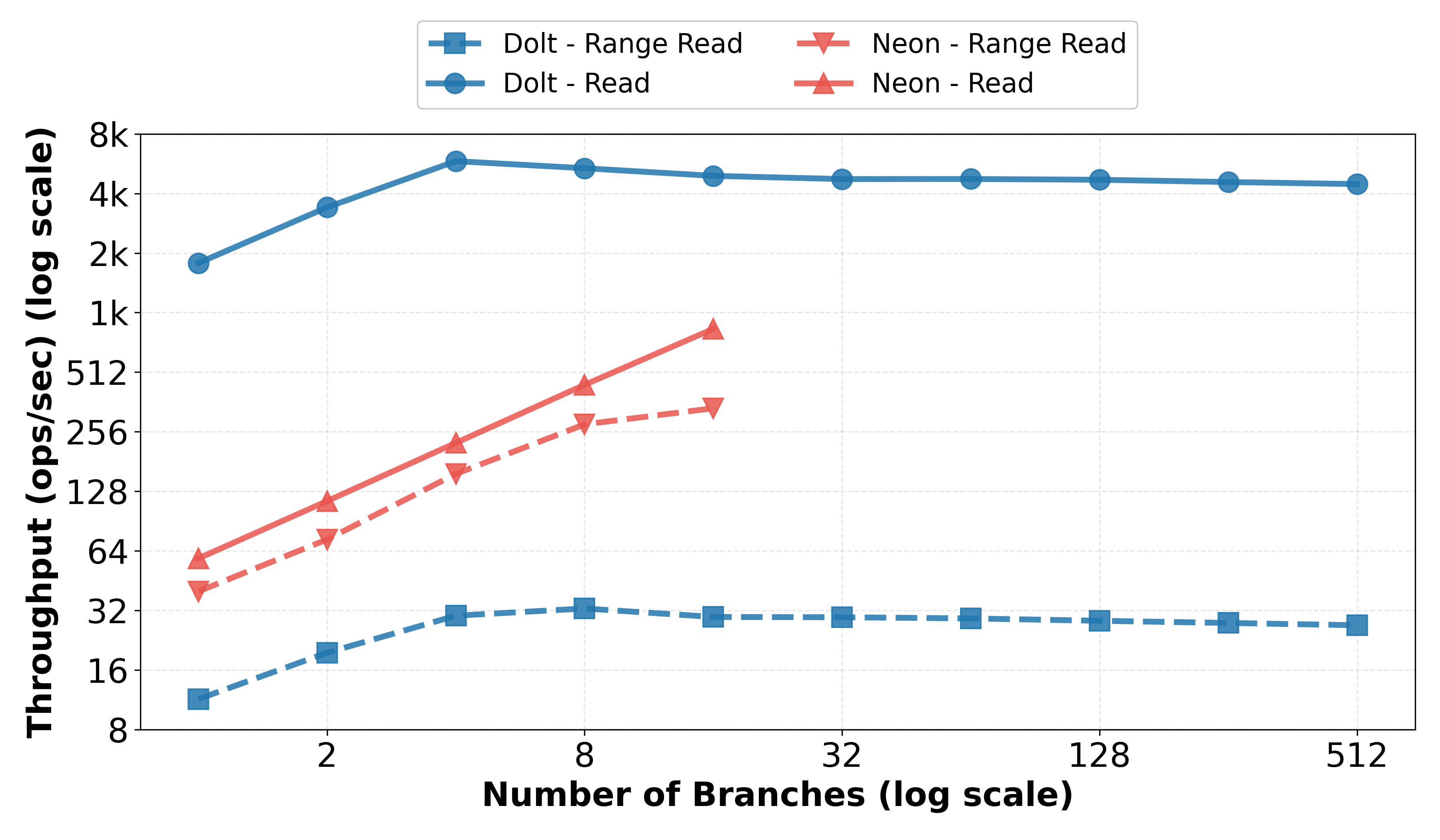}
    \caption{System reads capacity as number of branches increase (range size = 100).}
    \label{fig:tp-pptn}
\end{figure}

\stitle{Multi-Branch Scaling.}
Figure~\ref{fig:tp-pptn} reveals a fundamental architectural trade-off.
Neon's aggregate throughput scales near-linearly with the number of concurrent branches because each branch is provisioned with a dedicated compute instance, providing independent CPU and I/O resources.
This design trades higher per-branch provisioning cost for the ability to scale read capacity by adding branches---up to the 20 concurrent branch limit imposed by the cloud service.
In contrast, Dolt's throughput plateaus at 4 branches because all branches execute within a single process and share a fixed resource pool.
Adding branches provides no additional compute resources, and concurrent threads contend for CPU and disk I/O.
This design minimizes branch overhead and operates at fixed server cost but cannot scale read capacity beyond the single-server limit.
In summary, Neon's architecture trades high per-branch resource cost for parallel execution capacity, while Dolt trades limited scalability for low-overhead branching.
Consequently, Neon is cost-effective when expensive per-branch operations (e.g., range scans) amortize the provisioning overhead, whereas Dolt is cost-effective when branch creation frequency dominates and operations are lightweight (e.g., point lookups).

This analysis demonstrates the utility of our benchmark suite: the macrobenchmarks surfaced performance bottlenecks under realistic agentic workflows, while the microbenchmarks isolated the underlying architectural trade-offs that explain these bottlenecks.

\section{Related Work}
\label{sec:related}

%We position \sys ~~ relative to prior work on database branching mechanisms, database benchmarking methodologies, and version control systems for data management.

\subsection{Branching in OLAP systems}
\sys focuses on transaction-heavy agentic workloads that run on RDBMSes designed for highly mutable state, strict transactional guarantees, and low-latency operations.  
In contrast, analytical agentic workloads~\cite{liu2025supporting} operate over data lake systems optimized for large scans and bulk appends. These systems typically build on top of immutable object stores and implement branching via metadata operations: logical tables reference sets of physical objects, and writes materialize new objects while updating metadata (copy-on-write). Representative systems include:

\begin{itemize}
\item \textbf{LakeFS}~\cite{lakefs} provides repository-level Git-like operations (branch, commit, merge) over data lakes. It targets collaborative data engineering workflows.  

\item \textbf{Apache Iceberg}~\cite{apache_iceberg} uses table-level branch and tag primitives to support Write-Audit-Publish (WAP) workflows, where an isolated audit branch can be used to audit data quality before publishing to production.

\item \textbf{Delta Lake}~\cite{delta_lake} supports ACID semantics and time travel along a linear (single-branch) version history. The focus is on incremental and streaming pipelines rather than divergent branching.

\item \textbf{MotherDuck}~\cite{motherduck} supports zero-copy database clones for serverless analytics, and only materializes deltas between versions.  

\item \textbf{Bauplan}~\cite{bauplan} combines Git-like branching with isolated, branch-scoped execution environments. This enables versioned development contexts for experimentation of e.g., ETL pipelines.
\end{itemize}

\subsection{Database Benchmarking}  
\sys builds on past database benchmarks to create a parameterized, reproducible workload generator and evaluation suite.   We use data from CH-benCHmark~\cite{cole2011mixed}, which combines TPC-C and -H benchmarks to evaluate HTAP systems. YCSB~\cite{Cooper2010BenchmarkingCS} and OLTP-Bench~\cite{Difallah2013OLTPBenchAE} highlighted the value of workload parameterization and extensible benchmarking frameworks.

However, existing benchmarks assume a single-branch model that is not applicable to agentic exploration that demands tree-structured branch states. A single task may generate thousands of ephemeral branches, perform mutations in isolation, compare states across branches, and discard large subtrees. This introduces new workload dimensions that go beyond prior benchmarks, including branch creation and restoration, scalability to various branch tree-topologies, cross-branch query costs, schema and physical design operations, metadata overhead under high branch cardinality, and garbage collection efficiency for short-lived states.

To this end, \sys introduces a parameterized framework to model the {\it``branch-mutate-evaluate''} loop characteristic of agentic workloads, and supports microbenchmarks to measure individual primitive operations as well as macrobenchmarks motivated by future agentic data applications.

%However, all existing database benchmarks assume a linear state evolution model: transactions from concurrent users operate over a shared commit timeline. Performance metrics therefore focus on throughput, latency distributions, and concurrency control overhead under contention. Even HTAP benchmarks such as CH-benCHmark stress coexistence of reads and writes over a single logical database state.

% \input{sections/future_work.tex}
\section{Conclusion}
\label{sec:conclusion}

Agentic systems introduce a new database paradigm where even single agents generate thousands of speculative branches through high-frequency forking, evaluation, and pruning.
This paper presented \sys, the first systematic benchmark for evaluating database branching mechanisms under agentic workloads.
We characterized agentic branching through five representative workflows (software development, failure reproduction, data cleaning, MCTS, simulation), and developed a parametrized benchmark suite that instantiates the ``branch--mutate--evaluate--prune'' loop and periodic background cross-branch aggregation.

Evaluation of production systems reveals fundamental trade-offs in branch resource management: systems allocating dedicated compute instances per branch (e.g., Neon) incur higher branch creation and switching costs but provide better performance, whereas systems managing all branches within a single instance (e.g., Dolt) enable faster branching operations but their data operations suffer from higher overheads.
Additionally, existing systems lack native cross-branch aggregation support, a critical primitive for agentic workflows that must compare outcomes across speculative paths.
Supporting these emerging agentic exploration workloads demands elevating branch instantiation latency, metadata scalability, and cross-branch analytics to first-class design considerations.
% Future work includes multi-agent scenarios, lakehouse evaluation, and merge semantics characterization.

% Add a dummy citation to resolve citation warning
% Remove this line once you add actual citations in your sections
\nocite{*}

% \input{template.tex}

%\begin{acks}
%  This work was supported by the [...] Research Fund of [...] (Number [...]). Additional funding was provided by [...] and [...]. We also thank [...] for contributing [...].
%\end{acks}

%\clearpage

\bibliographystyle{ACM-Reference-Format}
\bibliography{ref}
\clearpage

\appendix

\section{Macrobenchmark Queries}\label{a:macroqs}

This section provides examples of the queries used in the macrobenchmark.

\subsection{Software Development}

\noindent
\fbox{%
\parbox{0.98\columnwidth}{%
\textbf{Schema Change:}\\
\texttt{ALTER TABLE customer ADD COLUMN loyalty\_tier\_xx;}\\[2pt]
\textbf{Data Mutation (backfill):}\\
\texttt{UPDATE customer SET loyalty\_tier\_xx = (CASE WHEN ...);} \\[2pt]
\textbf{Read (evaluate quality):}\\
\texttt{SELECT loyalty\_tier\_xx, COUNT(*) FROM customer GROUP BY loyalty\_tier\_xx;}\\[2pt]
\textbf{Compare (cross-branch):}\\
\texttt{SELECT loyalty\_tier\_xx, COUNT(*) AS tier\_count,}\\
\texttt{\quad AVG(c\_ytd\_payment) AS avg\_payment}\\
\texttt{FROM customer GROUP BY loyalty\_tier\_xx;}
}%
}

\subsection{Failure Reproduction}

\noindent
\fbox{%
\parbox{0.98\columnwidth}{%
\textbf{Schema Change (replay):}\\
\texttt{ALTER TABLE order\_line ADD COLUMN X;}\\
\texttt{ALTER TABLE orders ADD/DROP COLUMN Y;}\\
\texttt{ALTER TABLE customer ADD COLUMN Z;}\\[2pt]
\textbf{Data Mutation (replay):}\\
\texttt{INSERT INTO orders (...);}\\
\texttt{UPDATE customer SET (...);}\\
\texttt{DELETE FROM order\_line WHERE (...);}\\[2pt]
\textbf{Read (invariant check):}\\
\texttt{SELECT DISTINCT o.o\_id FROM orders}\\
\texttt{\quad JOIN order\_line ON (...)}\\
\texttt{\quad GROUP BY (...) HAVING (...);}
}%
}

\subsection{Data Cleaning}

\noindent
\fbox{%
\parbox{0.98\columnwidth}{%
\textbf{Schema Change (aux column):}\\
\texttt{ALTER TABLE customer ADD COLUMN c\_clean\_xx DEFAULT FALSE;}\\[2pt]
\textbf{Data Mutation (strategies):}\\
\texttt{UPDATE customer SET c\_balance = 0 WHERE c\_balance IS NULL;}\\
\texttt{DELETE FROM customer WHERE c\_balance IS NULL;}\\[2pt]
\textbf{Read (correctness):}\\
\texttt{SELECT COUNT(CASE WHEN c\_balance < 0 THEN 1 END)}\\
\texttt{\quad AS invalid FROM customer;}\\[2pt]
\textbf{Compare (cross-branch):}\\
\texttt{SELECT COUNT(...) AS invalid,}\\
\texttt{\quad MAX(c\_ytd\_payment)-MIN(c\_ytd\_payment) AS spread}\\
\texttt{FROM customer;}
}%
}

\subsection{Monte Carlo Tree Search}

\noindent
\fbox{%
\parbox{0.98\columnwidth}{%
\textbf{Data Mutation:}\\
\texttt{UPDATE stock SET s\_quantity = (...) WHERE (...);}\\[2pt]
\textbf{Read (expected reward):}\\
\texttt{SELECT SUM(ol\_amount) AS total\_cost}\\
\texttt{FROM order\_line JOIN warehouse ON (...);}
}%
}

\subsection{Simulation}

\noindent
\fbox{%
\parbox{0.98\columnwidth}{%
\textbf{Data Mutation:}\\
\texttt{INSERT INTO orders (...);}\\
\texttt{UPDATE stock SET (...) WHERE (...);}\\
\texttt{INSERT INTO order\_line (...);}\\[2pt]
\textbf{Read (evaluate outcomes):}\\
\texttt{SELECT SUM(...) AS stockouts,}\\
\texttt{\quad SUM(ol.ol\_amount) AS total\_cost}\\
\texttt{FROM stock JOIN order\_line ON (...);}\\[2pt]
\textbf{Compare (aggregate across branches):}\\
\texttt{SELECT AVG(stockouts), AVG(total\_cost)}\\
\texttt{FROM all\_branches.outcome\_metrics;}
}%
}

\end{document}